\documentclass[a4paper,10pt]{article}

\pdfoutput=1
\usepackage{amsmath}
\usepackage{amssymb}
\usepackage{jcappub}
\usepackage{mathtools}
\usepackage{graphicx}
\usepackage[dvipsnames]{xcolor}
\usepackage{footmisc}
\usepackage[normalem]{ulem}
\usepackage{float}
\usepackage{braket}
\usepackage{wrapfig}
\usepackage{lscape}
\usepackage{rotating}
\usepackage{epstopdf}
\usepackage{comment}
\usepackage[numbers, sort&compress]{natbib}
\usepackage{tikz, pgfplots}
\usetikzlibrary{positioning}
\usetikzlibrary{decorations.pathmorphing}
\usepackage[export]{adjustbox}[2011/08/13]

\tikzset{snake/.style={decorate, decoration=snake}}

\newcommand{\be}{\begin{equation}}
	\newcommand{\ee}{\end{equation}}

\begin{document}
	\title{Dark winds on the horizon: Prospects for detecting neutrino and hot dark matter wakes in large-scale structure}
	
	\author[1, 2]{Caio B. de S. Nascimento}	
	\emailAdd{caiobsn@perimeterinstitute.ca}
	\author[1]{and Marilena Loverde}
	\emailAdd{mloverde@uw.edu}
	
	\affiliation[1]{Department of Physics, University of Washington, 1410 NE Campus Pkwy, Seattle, USA}
	\affiliation[2]{Perimeter Institute for Theoretical Physics, 31 Caroline Street N., Waterloo, Ontario, Canada}

\abstract{We explore the cosmological signatures of neutrino and Hot Dark Matter (HDM) wakes, which refers to the preferential accumulation of neutrinos (or, more broadly, HDM particles) downstream of moving cold dark matter structures. We improve on existing theoretical models, and provide forecasts for the detectability  of the effect in future surveys under more realistic conditions than previously considered in the literature. We show that neutrino and HDM wakes are unlikely to be ever observed with the most natural tracer of a hot subcomponent of the total dark matter  on cosmological scales, i.e. 2D weak lensing surveys. However, the effect can be detected at a high significance with idealistic 3D maps of a tracer of HDM, for sufficiently small values of the effective free-streaming length (e.g. present-day values of $k_{\textrm{fs},0} \gtrsim 0.03\textrm{Mpc}^{-1}$ to reach $\textrm{SNR} \gtrsim 1$, for a HDM species accounting for a percent of the total dark matter). HDM wakes are a smoking gun of the effects of free-streaming, which cannot be mimicked by changes to the background expansion history (such as allowing for the dark energy to be dynamical), and hence offer another avenue to search for massive neutrinos, and hot subcomponents of the total dark matter more broadly, in a way that complements traditional observables. }

\maketitle

\section{Introduction}
\label{sec:int}

The field of Cosmology has entered a precision era where new (and also old, but yet undetected) ingredients pertaining the energy content of the universe, its initial conditions and gravitational evolution beyond General Relativity can be tested against cosmological observations. Many ongoing and upcoming surveys will collect unprecedented amounts of data, and it may well be that a major scientific discovery, one that will significantly change our understanding of the universe, will be made in upcoming decades \cite{SimonsObservatory:2025wwn, CMB-S4:2022ght, DESI:2025fxa, Ferraro:2022cmj, Euclid:2024yrr, crill2020spherex, Mao:2022fyx, Dore:2019pld, Gong:2025ecr}. 

One outstanding open question that near future astronomical surveys will help to address is the nature of dark matter. We can infer its existence and overall abundance from its gravitational effects, but we do not yet know its correct microphysical description \cite{Yu:2025rez, Cirelli:2024ssz}. The possibilities for what dark matter could be range from fuzzy ultralight fields to primordial black holes (in terms of the mass of the underlying degree of freedom). Whatever it may be, it should look like a cold collisionless species on cosmological scales in order to fit existing observations. The totality of dark matter can be warm as long as the associated Jeans instability scale is sufficiently small, and alternatively the Jeans scale of a subcomponent of dark matter can be as large as the Hubble scale, as long as its fractional contribution to the overall matter density is sufficiently small \cite{Peters:2023asu, Garcia-Gallego:2025kiw, Euclid:2024pwi, Dror:2024ibf, Xu:2021rwg, Dayal:2023nwi, Verdiani:2025jcf, Boyarsky:2008xj}. In this work we will be interested in the latter possibility of a thermal species (with an effective free-streaming scale of cosmological relevance) making a small contribution to the total dark matter abundance, which we refer to as a Hot Dark Matter (HDM) species \footnote{Our distinction between hot and Warm Dark Matter (WDM) scenarios is based on the size of their corresponding effective free-streaming scales, which is of cosmological relevance in the case of HDM, and of astrophysical extent for WDM.}.  

This scenario is exactly realized by massive neutrinos. We now know from neutrino oscillation experiments that at least two of the three neutrino eigenstates are massive, with a lower bound on the sum of the masses of $M_{\nu} =\sum_{\nu} m_{\nu} \geq 0.06\textrm{eV} \ (0.1\textrm{eV})$ in the normal (inverted) hierarchy \cite{deSalas:2020pgw, Esteban:2020cvm, Gonzalez-Garcia:2021dve}. An upper bound to a weighted mass average has been obtained with beta decay measurements made by the KATRIN experiment, $m_{\nu,\beta} < 0.45\textrm{eV}$ at $90\%$ confidence level \cite{KATRIN:2024cdt}. Much stronger upper bounds to the sum of the neutrino masses have been obtained from cosmological observations (under the assumption of the $\Lambda$CDM cosmology, and its variants) \cite{Planck:2018vyg, ACT:2025tim, DESI:2024hhd, Ivanov:2024jtl}. The recent measurement of Baryon Acoustic Oscillations (BAO) in the clustering of galaxies and quasars from DESI DR2 implies, in combination with Cosmic Microwave Background (CMB) observations from Planck, that $M_{\nu} < 0.0642 \textrm{eV}$ at $95\%$ confidence level \cite{DESI:2025ejh}. These existing cosmological constraints on the neutrino mass scale can be readily interpreted as bounds on the abundance and effective free-streaming scale of more general HDM models \cite{Green:2021xzn}. 

On one hand these impressive constraints on neutrino masses, and on HDM models more broadly, highlight the power of cosmological observations and their potential to discover new physics beyond the standard model. On the other hand, their interpretation requires great caution due to a number of subtleties. First of all, they derive from the assumption of a $\Lambda$CDM cosmology and can be significantly relaxed when new ingredients are introduced \cite{Escudero:2020ped, daFonseca:2023ury, FrancoAbellan:2021hdb}, such as allowing for an evolving dark energy equation of state \cite{Hannestad:2005gj, DESI:2025fii, Reboucas:2024smm} \footnote{The upper bounds on $M_{\nu}$ are, however, reasonably stable under the addition of many other ingredients to the vanilla $\Lambda$CDM model \cite{Racco:2024lbu}.}. In addition, the tight upper bounds on the neutrino mass scale $M_{\nu}$ (which are starting to show some tension with the lower bounds from neutrino oscillations) do not derive from a correspondingly high constraining power alone, but rather also reflect the preference of the data for a negative effective neutrino mass scale \cite{Green:2024xbb, Craig:2024tky}. Such a preference is intrinsically tied to the degeneracy between $M_{\nu}$ and background parameters such as the fractional contribution of matter to the total energy density, $\Omega_{\textrm{m},0}$ \cite{Loverde:2024nfi, Lynch:2025ine} (although excess lensing of the CMB is also a contributing factor \cite{Graham:2025fdt}). This means that a future robust cosmological measurement of $M_{\nu}$ will require exquisite control over the background expansion history. As such, it is not clear at the moment whether the preference for a negative effective neutrino mass is due to the manifestation of new unexpected physics or simply a diagnostic for the incompatibility between different datasets.

The points raised above offer additional motivation to explore potential new signatures of neutrino masses, and more broadly hot subcomponents of the total dark matter, in cosmological observations \cite{Tishue:2025zdw, LoVerde:2016ahu, Rogozenski:2023tse, Marques:2018ctl, Kreisch:2018var, Shiveshwarkar:2020jxr, Pal:2025hpl}. This work will focus on HDM wakes, building from the example of massive neutrinos. The HDM does not clump at scales below the effective free-streaming length, $k \gtrsim k_{\textrm{fs}}$, sourcing a nonzero relative velocity between hot and Cold Dark Matter (CDM) species. The (local) distribution of HDM then picks up a dipole along the direction of the relative motion, forming wakes downstream of moving CDM structure. While this signal is faint, it has the advantage of being completely independent from the background cosmology, in the sense that it is nonvanishing only in the presence of a relative velocity perturbation between CDM and a subdominant HDM species. This turns HDM wakes into a clear smoking gun of the effects of free-streaming, that is highly complementary to the traditional signal of power spectrum suppression on small scales \footnote{While HDM wakes effects cannot be mimicked by changes to the background expansion history, the relative velocity between hot and cold dark matter can in principle be imprinted on the initial conditions, rather than dynamically via free-streaming, generating similar signatures. This happens, for example, in the case of relative velocity perturbations between CDM and baryons \cite{tseliakhovich2010relative, mcquinn2012impact}. }. 

This signal was originally proposed in \cite{Zhu:2013tma}, and since then many other works have appeared forecasting the observability of the effect in future surveys \cite{Zhu:2019kzb, Zhu:2014qma, Okoli:2016vmd, Ge:2023nnh, Ge:2024kac, Ge:2025ctw}, and looking for it in simulations of large-scale structure formation \cite{Inman:2016prk, Hernandez-Molinero:2024sds}. Despite this significant activity, theory efforts are still required to properly model the physical effects of HDM wakes and to completely characterize the space of observables. These efforts also enable more realistic forecasts on the observability of HDM wakes in the large-scale structure \cite{Nascimento:2023ezc}. In this work we make significant progress in these directions.

We improve the theoretical modeling of HDM wakes, and build upon these new tools to show that measuring the effect will require two separate 3D maps of both CDM and HDM tracers (or at least two independent tracers that depend on different linear combinations of CDM and HDM). A very natural tracer of HDM is weak gravitational lensing, which is sensitive to the total matter field. However, due to its 2D nature, it is unlikely that HDM wakes will ever be observed with this probe as we will explain in Sec.~\ref{sec:2d}. The typical example of a 3D galaxy survey does not work either, as galaxies form within halos and hence are expected to mostly trace cold dark matter and baryons, and not to (directly) trace a sub-component of the total dark matter with a free-streaming scale of cosmological relevance (which is homogeneously distributed on halo scales) \footnote{This intuition was precisely confirmed with simulations in the case of massive neutrinos, and it is usually refereed to as the ``cb prescription" \cite{LoVerde:2014pxa, Villaescusa-Navarro:2017mfx, Castorina:2015bma, Costanzi:2013bha, Castorina:2013wga, Villaescusa-Navarro:2013pva}. \label{foot_2}}. Similarly, the quadrupole distortion in the distribution of galaxies due to redshift-space distortions is expected to trace the velocity field of cold dark matter and baryons alone \cite{Verdiani:2025znc, marulli2011effects}. As a result, there are no natural observables to detect HDM wakes in the near future, although the effect is in principle observable as we show in Sec.~\ref{sec:3d}. 

The paper is organized as follows. In Sec.~\ref{sec:wakes} we detail our theoretical modeling of HDM wakes, based on the specific example of massive neutrinos, highlighting the improvements made over previous works. In Sec.~\ref{sec:detection} we work out detection prospects for HDM wakes based on quadratic estimators of relative velocity reconstruction, building upon our improved theoretical model. We summarize our results in Sec.~\ref{sec:concl}. Finally, three appendices complement the main text with auxiliary calculations. Appendix \ref{sec:app1} contains a derivation of the short-scale ($k \gg k_{\textrm{fs}}$) limit of the neutrino dipole,  Appendix \ref{sec:app2} details the derivation of cumulative signal-to-noise ratios based on the unbiased optimal quadratic estimators employed, and Appendix \ref{sec:app3} works out the optimal combination of redshift bins to maximize sample variance cancellation and the signal-to-noise, in the two-dimensional case.

\section{Theoretical modeling of HDM wakes}
\label{sec:wakes}

We will build our general theory model of HDM wakes from the specific example of massive neutrinos. In that case, both the overall abundance of a nonrelativistic massive neutrino species (at late-times) and its associated free-streaming scale are fully specified by the neutrino mass, $m_{\nu}$ \cite{Lesgourgues:2006nd}. The fractional contribution of the neutrino species to the total matter energy density is \footnote{We adopt the subscript h, instead of the usual $\nu$, because our goal is to consider a broader scope of HDM models. The massive neutrino case then follows as an example.}
\begin{equation}
\label{eq:mnu_abundance}
	f_{\textrm{h}} = \frac{m_{\nu}}{93.14 \textrm{eV} h^2 \Omega_{\textrm{m}}} \,,
\end{equation}
where $H_{0} = 100h \ \textrm{km/s/Mpc}$ is the present-day value of the Hubble expansion rate and $\Omega_{\textrm{m}}$ the present-day fractional contribution of matter (including massive neutrinos) to the critical energy density. For numerical calculations in this paper, we adopt a fiducial flat $\Lambda$CDM cosmology with $h=0.7$ and $\Omega_{\textrm{m}}=0.3$. The neutrino free-streaming scale is given by \cite{Ringwald:2004np, ali2013efficient}
\begin{equation}
\label{eq:free_streaming}
	k_{\textrm{fs}} = \sqrt{\frac{\log 4}{2\zeta(3)} \Omega_{\textrm{m}}} \frac{m_{\nu}H_{0}}{T_{\nu,0}} \sqrt{a} \equiv k_{\textrm{fs},0}\sqrt{a} \,,
\end{equation}
where $T_{\nu,0} \approx 1.95\textrm{K} \approx 1.7 \times 10^{-4} \textrm{eV}$ is the neutrino temperature today, and we assume neutrinos are nonrelativistic. Our effective model of HDM will then follow from keeping both $f_{\textrm{h}}$ and $k_{\textrm{fs},0}$ as free parameters (while assuming the same redshift dependence of the effective free-streaming scale, $k_{\textrm{fs}} \propto \sqrt{a})$. To accomplish this, we will introduce approximate formulas for the ratio of neutrino and CDM transfer functions that depend only on the neutrino free-streaming scale, and then apply these formulas for arbitrary $k_{\textrm{fs},0}$. Of course, such an approach relies on the assumption of similarity between the HDM and standard neutrinos properties, and breaks down for more complex models that are not neutrino-like.

We will work to leading order in $f_{\textrm{h}} \ll 1$, and within the approximation of a linear response of massive neutrino perturbations to the fully nonlinear distribution of CDM. The former condition guarantees that any gravitational backreactions can be neglected when computing the neutrino density contrast $\delta_{\textrm{h}}(\eta,\vec{x})$, which is sourced solely by the CDM density contrast $\delta_{\textrm{c}}(\eta,\vec{x})$ \footnote{We include baryons in the definition of CDM as well, while ignoring collision terms due to baryonic effects. \label{foot_1}}. The latter condition enables us to write the following simple solution to the collisionless Boltzmann (or Vlasov) equation \cite{Ringwald:2004np, ali2013efficient, Holm:2024zpr, Hotinli:2023scz}
\begin{equation}
\label{eq:vlasov_sol}
	\delta_{\textrm{h}}(\eta,\vec{k}) = \frac{3}{2} \Omega_{\textrm{m},0} H_{0}^{2} \int_{0}^{\eta} d\eta' a(\eta') (\eta-\eta') \Delta\left(k\frac{T_{\nu,0}}{m_{\nu}}(\eta-\eta')\right) \delta_{\textrm{c}}(\eta',\vec{k}) \,,
\end{equation} 
where we work in Fourier space, with the superconformal time $\eta$, defined in terms of the cosmic time $t$, via $dt=a^{2}d\eta$. The function $\Delta(x)$ is defined as
\begin{equation}
\label{eq:aux_function}
	\Delta(x) \equiv \int_{0}^{\infty} \frac{y^2 dy}{3 \zeta(3)} f_{\textrm{FD}}(y) j_{0}(xy) \,,
\end{equation} 
with $f_{\textrm{FD}}(y)=2/(1+e^{y})$ the Fermi-Dirac profile and $j_{0}(x) = \sin x/x$ the spherical Bessel function of order zero. Specific HDM models will depend on various different ingredients that deviate from our assumptions, which are tailored to the case of a standard neutrino species. For instance, the background distribution function can easily deviate from a Fermi-Dirac profile. We expect these model dependencies to make order unity changes to our signal-to-noise predictions, with the neutrino-specific ingredients we adopt providing a robust order of magnitude estimation. 

In \cite{Nascimento:2023ezc} we derived the neutrino dipole distortion due to CDM bulk flows from first principles, starting from Eq.~(\ref{eq:vlasov_sol}). We were able to do so in the limit $k \gg k_{\textrm{fs}}$, and under the assumption that $k_{\textrm{fs}} \ll k_{\textrm{NL}}$, where $k_{\textrm{NL}} \sim 0.1\textrm{Mpc}^{-1}$ is the scale associated to the bulk flow of CDM, also usually known in the literature as the scale of nonlinearities \footnote{The scale of nonlinearities can be defined in many different ways, all of which produce a present-day value of about $k_{\textrm{NL}} \sim 0.1\textrm{Mpc}^{-1}$.}. This precise calculation allowed us to identify the following bispectrum \footnote{We follow the standard convention that the primed correlator has a factor of $(2\pi)^{3} \delta^{(3)}(\vec{k}+\vec{k'}-\vec{K})$ stripped of. We also omit time dependencies for simplicity of notation.}
\begin{equation}
\label{eq:bispectrum}
B(k,k',K) = \frac{1}{2} \left\langle \left[\delta_{\textrm{h}}(\vec{k}) \delta_{\textrm{c}}(\vec{k}') - \delta_{\textrm{c}}(\vec{k}) \delta_{\textrm{h}}(\vec{k}')\right] \phi_{\textrm{rel}}(-\vec{K}) \right\rangle^{'} \,,
\end{equation}
as the lowest order correlation function which is a smoking gun of HDM wakes, and can potentially be detected with future surveys. Here $\phi_{\textrm{rel}}$ is a scalar potential to the relative velocity field $\vec{v}_{\textrm{rel}} = \vec{v}_{\textrm{c}}-\vec{v}_{\textrm{h}}$, defined by $\vec{v}_{\textrm{rel}}(\vec{K}) = i\hat{K} \phi_{\textrm{rel}}(\vec{K})$ in Fourier space. 

The observable in Eq.~(\ref{eq:bispectrum}) has a straightforward physical interpretation. At physical scales below the relative velocity coherence length, $k \gtrsim k_{\textrm{coh}}$, the local CDM-HDM cross correlation picks up a dipole along the direction of the (approximately) uniform relative velocity, which manifests itself as in imaginary contribution to the local cross power spectrum. This term vanishes upon taking the global average since different relative velocity coherence regions have relative velocity fields pointing in different arbitrary directions. Hence, in order to extract the signal one needs to multiply the local dipole by the local relative velocity before taking the average, and Eq.~(\ref{eq:bispectrum}) is nothing but the Fourier space manifestation of this procedure (see \cite{Inman:2016prk} for a complementary real space approach to this observable from simulations). Note that on sufficiently large scales $\phi_{\textrm{rel}}(\vec{K}) \propto \delta_{\textrm{c}}(\vec{K})$ via the continuity equation, such that one could replace $\phi_{\textrm{rel}}(-\vec{K})$ with $\delta_{\textrm{c}}(-\vec{K})$ in the definition of Eq.~(\ref{eq:bispectrum}). However, the choice we made best aligns with the physical interpretation provided above.  

Many previous works have forecasted the detectability of neutrino wakes \cite{Zhu:2013tma, Zhu:2019kzb, Zhu:2014qma, Okoli:2016vmd, Ge:2023nnh, Ge:2024kac, Ge:2025ctw}, but most of them assume that galaxies directly trace the hot dark matter species with order unity bias, or that redshift space distortions trace the total matter field (including contributions from the HDM), which are likely not realistic assumptions as we argue later. Additionally, most of these previous works model wakes effects in terms of an imaginary contribution to the CDM-$\nu$ cross power spectrum, which can hide some of the underlying assumptions involved in the calculations. Instead, the more accurate approach is to model the signal in terms of a bispectrum, as in Eq.~(\ref{eq:bispectrum}). This was first realized by the authors of \cite{Zhu:2019kzb}, which also introduced the idea of rephrasing the measurement of neutrino wakes in terms of a reconstruction of the relative velocity field. This opens up the possibility of employing well-studied quadratic estimator techniques which were previously applied in the context of CMB lensing \cite{Hu:2001kj, Hirata:2003ka, Maniyar:2021msb} and KSZ tomography \cite{Smith:2018bpn, Deutsch:2017ybc, Hotinli:2025tul}. However, their quantitative results are inaccurate as they work within a purely kinematic treatment of the effect, which is in fact dynamical as can be seen from the observation that one needs to solve the Vlasov equation to properly derive it. 

We were the first to forecast the observability of neutrino wakes within a fully self-consistent theoretical framework based on the bispectrum \cite{Nascimento:2023ezc}. However, the approach taken in that work has some already mentioned limitations, namely it assumes $k_{\textrm{fs}} \ll k_{\textrm{NL}}$, and $k \gg k_{\textrm{fs}}$ for the comoving scales probed. On top of that, our original simplified forecast assumed a noiseless 3D map of the total matter field, and did not account for the complications associated with a realistic 2D tomographic survey. 

In this work we will fill in these gaps. On the theoretical modeling side, we rephrase the calculations in terms of a general HDM model, extending them to arbitrary values of $k_{\textrm{fs}}$, and also dropping the assumption that the physical scales probed are smaller than the effective free-streaming scale. To accomplish this we will assume that the squeezed limit, with $K \ll k$, is implied in Eq.~(\ref{eq:bispectrum}), such that we can use separate universe techniques. This approach has been taken in many previous works on the subject \cite{Zhu:2013tma, Zhu:2014qma, Okoli:2016vmd, Ge:2023nnh, Ge:2024kac, Inman:2016prk, Ge:2025ctw}, but not yet within the context of the full bispectrum description based on Eq.~(\ref{eq:bispectrum}). One important caveat to this proposal is that, in reality, we found in \cite{Nascimento:2023ezc} that the signal-to-noise ratio peaks at a nonzero $K/k$ \footnote{There are two reasons for this. First, the relative velocity potential $\phi_{\textrm{rel}}(\vec{K})$ dies away as $K \to 0$ which suppresses the squeezed limit in the signal. Second, the number of modes is suppressed in the squeezed limit as well, which pushes the cumulative signal-to-noise peak away from $K/k \to 0$.}, so our results should be interpreted with care, as an order of magnitude estimation of the signal-to-noise associated to HDM wakes \footnote{Should it ever be the case that this signal can be realistically detected in real data, we would potentially need to build a much more sophisticated theoretical framework that can handle triangle configurations away from the squeezed limit, while also avoiding the aforementioned intrinsic limitations to the approach taken in \cite{Nascimento:2023ezc}.}.  On the forecasting side we carefully evaluate what are the penalties, in terms of signal-to-noise ratio loss, associated to the more realistic case of noisy 2D maps of a HDM tracer. 

We are now ready to dive into the details of our theoretical model. Within the assumption of an implied squeezed limit, we can use the separate universe technique to compute the bispectrum. The first step is to add a background relative bulk flow between CDM and HDM
\begin{equation}
\label{eq:displacement}
	\vec{\psi}_{\textrm{rel}}(\eta) =  \int_{0}^{\eta} d\eta' a(\eta') \vec{v}_{\textrm{rel}}(\eta') \,,
\end{equation}
that can be absorbed into a coordinate transformation, $\vec{x} \to \hat{\vec{x}} = \vec{x} + \vec{\psi}_{\textrm{rel}}(\eta)$, such that the displaced CDM density profile reads
\begin{equation}
\label{eq:displaced_profile}
	\hat{\delta}_{\textrm{c}}(\eta,\vec{x}) = \delta\left(\eta, \vec{x}+ \vec{\psi}_{\textrm{rel}}(\eta)\right) \implies \hat{\delta}_{\textrm{c}}(\eta,\vec{k}) = e^{i\vec{k} \cdot \vec{\psi}_{\textrm{rel}}(\eta)} \delta_{\textrm{c}}(\eta,\vec{k}) \,.
\end{equation}   

Next we substitute Eq.~(\ref{eq:displaced_profile}) into Eq.~(\ref{eq:vlasov_sol}) to obtain an expression for $\hat{\delta}_{\textrm{h}}(\eta,\vec{x})$, and compute the CDM-HDM cross power spectrum in the presence of the long-wavelength mode $\vec{\psi}_{\textrm{rel}}(\eta)$. We find after a straightforward calculation 
\begin{equation}
\label{eq:cross_ps_longmode}
	\hat{P}_{\textrm{ch}}(\eta,\vec{k}) = \frac{3}{2} \Omega_{\textrm{m}} H_{0}^{2} \int_{0}^{\eta} d\eta' a(\eta') \frac{D_{\textrm{L}}(\eta')}{D_{\textrm{L}}(\eta)} (\eta-\eta') e^{-i\vec{k} \cdot \Delta \vec{\psi}_{\textrm{rel}}(\eta,\eta')} \Delta\left(k\frac{T_{\nu,0}}{m_{\nu}}(\eta-\eta')\right) P_{\textrm{cc}}(\eta,k) \,,
\end{equation}
where for simplicity we assume a linear theory evolution for the CDM density, $\delta_{\textrm{c}}(\eta,k) \propto D_{\textrm{L}}(\eta)$, with $D_{\textrm{L}}(\eta)$ the linear growth factor \footnote{However, we do make use of the fully nonlinear power spectrum for CDM, evaluated at $z=0$ with the Boltzmann solver CLASS via the HMcode \cite{Mead:2020vgs}. We do not expect the approximation of linear evolution with redshift (which overestimates the nonlinear power spectrum at earlier times) to dominate the error budget in our theoretical model, given the underlying assumption of the squeezed limit discussed in the main text.}. We also introduced the notation $\Delta \vec{\psi}_{\textrm{rel}}(\eta,\eta') = \vec{\psi}_{\textrm{rel}}(\eta) - \vec{\psi}_{\textrm{rel}}(\eta')$. Let us now consider a Taylor series expansion of the exponential in Eq.~(\ref{eq:cross_ps_longmode}). The zeroth-order term, which we call the monopole, is simply the standard CDM-HDM cross power spectrum. That is
\begin{equation}
	\label{eq:cross_ps_monopole}
	\hat{P}^{\textrm{mon}}_{\textrm{ch}}(\eta,k) = P_{\textrm{ch}}(\eta,k) = \frac{3}{2} \Omega_{\textrm{m}} H_{0}^{2} \int_{0}^{\eta} d\eta' a(\eta') \frac{D_{\textrm{L}}(\eta')}{D_{\textrm{L}}(\eta)} (\eta-\eta') \Delta\left(k\frac{T_{\nu,0}}{m_{\nu}}(\eta-\eta')\right) P_{\textrm{cc}}(\eta,k) \,.
\end{equation}
The first order term is the dipole, and it reads
\begin{equation}
	\label{eq:cross_ps_dipole}
	\hat{P}^{\textrm{dip}}_{\textrm{ch}}(\eta,\vec{k}) = - \frac{3}{2} \Omega_{\textrm{m}} H_{0}^{2} \int_{0}^{\eta} d\eta' a(\eta') \frac{D_{\textrm{L}}(\eta')}{D_{\textrm{L}}(\eta)} (\eta-\eta') i\vec{k} \cdot \Delta \vec{\psi}_{\textrm{rel}}(\eta,\eta') \Delta\left(k\frac{T_{\nu,0}}{m_{\nu}}(\eta-\eta')\right) P_{\textrm{cc}}(\eta,k) \,.
\end{equation}
In order to make contact with previous results in the literature, we want to extract the $k \gg k_{\textrm{fs}}$ limits of Eqs.~(\ref{eq:cross_ps_monopole}) and (\ref{eq:cross_ps_dipole}). Here we simply show the following final expression \footnote{Moving forward we will often omit time dependencies for simplicity of notation.}
\begin{equation}
\label{eq:short_scale_dipole}
	\hat{P}^{\textrm{dip}}_{\textrm{ch}}(\vec{k}) \underset{k\gg k_{\textrm{fs}}}{=} -\gamma \frac{i\vec{k}}{k} \cdot \frac{\vec{v}_{\textrm{rel}}}{\sigma_{\nu}} P_{\textrm{ch}}(k) \,,
\end{equation}
where
\begin{equation}
\label{eq:nu_velocity}
	\sigma_{\nu}(a) = \sqrt{\frac{3\zeta(3)}{\log 4}} \frac{T_{\nu,0}}{m_{\nu}a} \,,
\end{equation}
is the neutrino (thermal) velocity dispersion, and $\gamma = (\pi/6\zeta(3))(3\zeta(3)/\log 4)^{3/2} \approx 1.83$. Note that the dipole in Eq.~(\ref{eq:short_scale_dipole}) scales like $\sim v_{\textrm{rel}}/\sigma_{\nu}$. 

A detailed derivation of Eq.~(\ref{eq:short_scale_dipole}) is provided in Appendix \ref{sec:app1}. This exactly reproduces our previous results from \cite{Nascimento:2023ezc} in the $k \gg k_{\textrm{fs}}$ limit. One important aspect of Eq.~(\ref{eq:short_scale_dipole}) is that the dipole is proportional to the monopole, which directly implies two attractive features of HDM wakes from the observational point of view. First, it means that we do not actually need a theoretical model for the monopole in order to make a prediction for what the dipole should be. Instead, the monopole can be (in principle) measured directly and then used to inform Eq.~(\ref{eq:short_scale_dipole}). Second, the relative velocity can be reconstructed from a measurement of the ratio $\sim \hat{P}^{\textrm{dip}}_{\textrm{ch}}(\vec{k})/\hat{P}^{\textrm{mon}}_{\textrm{ch}}(k)$, which is independent of the specific realization of the primordial fluctuations, rendering the measurement to be cosmic variance free. This is what makes this faint signal potentially observable with futuristic experiments. 

We expect the dipole to die off on scales $k \ll k_{\textrm{fs}}$, since then the HDM behaves like CDM and there is no relative bulk motion between them. Within the approach in  \cite{Nascimento:2023ezc} we could not derive the detailed form of how the dipole approaches zero at sufficiently large scales, so instead we simply multiplied Eq.~(\ref{eq:short_scale_dipole}) by a phenomenological factor of $T_{\textrm{dip}}(k) = (1+k_{\textrm{fs}}/k)^{-1}$. Since we now have access to the full scale dependencies from Eqs.~(\ref{eq:cross_ps_monopole}) and (\ref{eq:cross_ps_dipole}), we can derive the exact shape of $T_{\textrm{dip}}(k)$. In preparation for this, we first use Eq.~(\ref{eq:displacement}) to note that
\begin{equation}
\label{eq:delta_displacement}
	\Delta \vec{\psi}_{\textrm{rel}}(\eta,\eta') = \vec{\psi}_{\textrm{rel}}(\eta) - \vec{\psi}_{\textrm{rel}}(\eta') = \int_{\eta'}^{\eta} d\eta'' a(\eta'') \vec{v}_{\textrm{rel}}(\eta'') = \vec{v}_{\textrm{rel}}(\eta) \int_{\eta'}^{\eta} d\eta'' a(\eta'') \frac{\sigma_{\textrm{rel}}(\eta'')}{\sigma_{\textrm{rel}}(\eta)} \,,
\end{equation}
where $\sigma_{\textrm{rel}} = \sqrt{\langle v_{\textrm{rel}}^2 \rangle}$ is the root-mean-square relative velocity, which shares the same time dependence as $v_{\textrm{rel}}$ itself. With this in mind, we want to generalize Eq.~(\ref{eq:short_scale_dipole}) to all scales. For this we consider an ansatz of the form
\begin{equation}
	\label{eq:dipole_all_scales}
	\hat{P}^{\textrm{dip}}_{\textrm{ch}}(\vec{k}) = -\gamma T_{\textrm{dip}}(k) \frac{i\vec{k}}{k} \cdot \frac{\vec{v}_{\textrm{rel}}}{\sigma_{\nu}} P_{\textrm{ch}}(k) \,.
\end{equation}
We may now combine Eqs.~(\ref{eq:cross_ps_monopole}), (\ref{eq:cross_ps_dipole}), (\ref{eq:delta_displacement}) and (\ref{eq:dipole_all_scales}) to arrive at
\begin{equation}
\label{eq:aux_function_2}
	T_{\textrm{dip}}(\eta,k) = k \frac{ \sigma_{\nu}}{\gamma} \frac{\int_{0}^{\eta} d\eta' a(\eta') D_{\textrm{L}}(\eta') (\eta-\eta') \Delta\left(k \frac{T_{\nu,0}}{m_{\nu}}(\eta-\eta')\right) \int_{\eta'}^{\eta} d\eta'' a(\eta'') \frac{\sigma_{\textrm{rel}}(\eta'')}{\sigma_{\textrm{rel}}(\eta)}}{\int_{0}^{\eta} d\eta' a(\eta') D_{\textrm{L}}(\eta') (\eta-\eta') \Delta\left(k \frac{T_{\nu,0}}{m_{\nu}}(\eta-\eta')\right)} \,.
\end{equation}
We have evaluated Eq.~(\ref{eq:aux_function_2}) numerically for different values of the neutrino mass and at different redshifts, extracting the relative velocity transfer functions from the Cosmic Linear Anisotropy Solving System (CLASS) \cite{blas2011cosmic}. We found the following numerical fit to work well
\begin{equation}
\label{eq:fit_dipole}
	T_{\textrm{dip}}(k) \approx \left[1+2.4\left(\frac{k_{\textrm{fs}}}{k}\right)^{0.8}\right]^{-1} \,,
\end{equation} 
where the time dependence is encapsulated by the scaling with scale factor seen in Eq.~(\ref{eq:free_streaming}) for the neutrino free-streaming scale. Note that $T_{\textrm{dip}} \to 0$ when $k\ll k_{\textrm{fs}}$, and $T_{\textrm{dip}} \to 1$ when $k\gg k_{\textrm{fs}}$ as expected. We also find two other numerical fits to ratios of neutrino to CDM transfer functions. The first is
\begin{equation}
\label{eq:ratio_densities}
	T_{\delta}(k) = \frac{\delta_{\textrm{c}}(\vec{k}) - \delta_{\textrm{h}}(\vec{k})}{\delta_{\textrm{c}}(\vec{k})} \approx \left[1+0.25\left(\frac{k_{\textrm{fs}}}{k}\right)^{1.3}\right]^{-1} \,,
\end{equation}
and the second reads
\begin{equation}
\label{eq:ratio_velocities}
	T_{\theta}(k) = \frac{\theta_{\textrm{c}}(\vec{k}) - \theta_{\textrm{h}}(\vec{k})}{\theta_{\textrm{c}}(\vec{k})} \approx \left[1+0.5\left(\frac{k_{\textrm{fs}}}{k}\right)^{1.5}\right]^{-1} \,,
\end{equation}
where $\theta_{A}(\eta, \vec{x}) = \vec{\nabla} \cdot \vec{v}_{A}(\eta, \vec{x})$ stands for the divergence of the velocity field, with $A=\textrm{c},\textrm{h}$. Eqs.~(\ref{eq:fit_dipole}), (\ref{eq:ratio_densities}) and (\ref{eq:ratio_velocities}) may now be evaluated within the context of a general HDM model where the present-day value of the free-streaming scale, $k_{\textrm{fs},0}$ in Eq.~(\ref{eq:free_streaming}), is taken to be a free parameter alongside the fractional contribution of hot to the total matter density, $f_{\textrm{h}}$.

To complete the theoretical model, we rewrite Eq.~(\ref{eq:dipole_all_scales}) in terms of the relative velocity scalar potential $\phi_{\textrm{ref}}(\vec{K})$, defined by $\vec{v}_{\textrm{rel}}(\vec{K}) = i\hat{K} \phi_{\textrm{rel}}(\vec{K})$ with $\vec{K}$ a long-wavelength mode ($K \ll k$). We arrive at
\begin{equation}
\label{eq:pre_squeezed_bispectrum}
	\langle \delta_{\textrm{h}}(\vec{k})\delta_{\textrm{c}}(\vec{k}') \rangle_{\textrm{dip}} = \frac{\gamma}{\sigma_{\nu}} T_{\textrm{dip}}(k) (\hat{k} \cdot \hat{K}) \phi_{\textrm{rel}}(\vec{K}) P_{\textrm{ch}}(k) \,,
\end{equation}   
where $\vec{K} = \vec{k}+\vec{k'}$. In what follows we will often work directly in terms of Eq.~(\ref{eq:pre_squeezed_bispectrum}), which fully characterizes the bispectrum in Eq.~(\ref{eq:bispectrum}). However, we also compute it directly for completeness
\begin{equation}
\label{eq:bispectrum_calculated}
	B(\vec{k},\vec{K}) = \left\langle \langle \delta_{\textrm{h}}(\vec{k})\delta_{\textrm{c}}(\vec{k}') \rangle_{\textrm{dip}} \phi_{\textrm{rel}}(-\vec{K}) \right\rangle^{'} = \frac{\gamma}{\sigma_{\nu}} T_{\textrm{dip}}(k) (\hat{k} \cdot \hat{K}) P_{\textrm{ch}}(k) P_{\phi_{\textrm{rel}} \phi_{\textrm{rel}}} (K) \,.
\end{equation}

The fit in Eq.~(\ref{eq:ratio_densities}) is used to compute the cross power spectrum $P_{\textrm{ch}}(k)$ from the CDM auto power spectrum, and Eq.~(\ref{eq:ratio_velocities}) is used to compute the power spectrum of the relative velocity potential $P_{\phi_{\textrm{rel}}\phi_{\textrm{rel}}}(K)$ from the CDM velocity power spectrum. Let us reinforce that we will evaluate Eq.~(\ref{eq:bispectrum_calculated}) for all possible values of $k$, $K$ and $\mu=\hat{k} \cdot \hat{K}$, although our derivation implicitly assumes the squeezed limit with $K \ll k$. For this reason, our results should be cautiously interpreted as an estimate of the size of HDM wakes effects. A complete characterization of the full bispectrum that works on all scales and for all values of the free-streaming length will require a significantly more sophisticated theoretical model (or simulations of nonlinear structure formation), which is beyond the scope of this work, but necessary for an accurate comparison of theoretical calculations with futuristic datasets. With this in mind, we are ready to explore detection prospects for the dipole induced by the relative bulk flow between CDM and a subdominant HDM species. 

\section{Detection prospects}
\label{sec:detection}

Our forecasts will be based on the basic idea that the relative velocity between HDM and CDM can be reconstructed from a measurement of the dipole, Eq.~(\ref{eq:pre_squeezed_bispectrum}) [see comments below Eq.~(\ref{eq:nu_velocity})]. For this we follow the approach first introduced in this context by the authors of \cite{Zhu:2019kzb}, with an improved theoretical model. We employ optimal unbiased quadratic estimators, which are very well-studied due to their extensive applications to CMB lensing \cite{Hu:2001kj, Hirata:2003ka, Maniyar:2021msb} and KSZ tomography \cite{Smith:2018bpn, Deutsch:2017ybc, Hotinli:2025tul}. 

Apart from the refined theoretical modeling discussed in Sec.~\ref{sec:wakes}, our work improves on the forecasts available in the literature in two additional fronts. First, the effect is rephrased in terms of a general HDM model where the present-day values of the abundance and free-streaming scales of a massive neutrino species are uncoupled, and are left as free parameters. Second, we account for projection effects and noise as appropriate to a 2D tomographic survey, as weak lensing is the most natural (direct) tracer of the HDM on cosmological scales.

\subsection{Detectability with 3D maps of HDM}
\label{sec:3d}

Let us begin with the idealistic scenario of a 3D tracer of HDM (in combination with a 3D galaxy survey), which for concreteness we take to be the total matter field. Following standard techniques, we write an estimator for the relative velocity potential of the following form
\begin{equation}
\label{eq:3d_estimator}
\begin{split}
	 \hat{\phi}_{\textrm{rel}}(\vec{K}) & = N(K) \int \frac{d^3\vec{k}}{(2\pi)^{3}} W(\vec{k},\vec{k'}) \frac{1}{2} \left[\delta_{\textrm{m}}(\vec{k})\delta_{\textrm{g}}(\vec{k}')-\delta_{\textrm{g}}(\vec{k})\delta_{\textrm{m}}(\vec{k}')\right] \\ & = N(K) \int \frac{d^3\vec{k}}{(2\pi)^{3}} W_{A}(\vec{k},\vec{k'}) \delta_{\textrm{m}}(\vec{k})\delta_{\textrm{g}}(\vec{k}') \,,
\end{split}
\end{equation} 
where in the second line of Eq.~(\ref{eq:3d_estimator}), $W_{A}(\vec{k},\vec{k}') = [W(\vec{k},\vec{k}')-W(\vec{k}',\vec{k})]/2$ stands for the antisymmetric combination of the filter function. Here $\delta_{\textrm{m}}=(1-f_{\textrm{h}}) \delta_{\textrm{c}}+f_{\textrm{h}} \delta_{\textrm{h}}$ is the density contrast of the total matter field, and $\delta_{\textrm{g}}=b\delta_{\textrm{c}}$ is the density contrast of the galaxy sample. Note that we assume the simplest possible model of galaxy clustering, based only on the linear galaxy bias $b$ \cite{Desjacques:2016bnm}. We also assume that galaxies only trace CDM (and baryons, see footnote \ref{foot_1}), but not the HDM (see footnote \ref{foot_2}). 

The normalization factor $N(K)$, and filter function $W(\vec{k},\vec{k}')$, derive from two constraints to be imposed on Eq.~(\ref{eq:3d_estimator}). The first is that it should be an unbiased estimator, namely $\langle  \hat{\phi}_{\textrm{rel}}(\vec{K}) \rangle = \phi_{\textrm{rel}}(\vec{K})$. Second is that we want it to be optimal, in the sense that it minimizes the covariance. We explain how these requirements uniquely fix the covariance of the estimator in Appendix \ref{sec:app2}, producing the following final formula
\begin{equation}
\label{eq:3d_covariance}
	N(K)^{-1} = \frac{1}{3} f_{h}^2 b^{2} \frac{\gamma^{2}}{\sigma_{\nu}^{2}} \int \frac{k^{2}dk}{2\pi^{2}} \frac{\left[T_{\textrm{dip}}(k)P_{\textrm{ch}}(k)\right]^2}{P_{\textrm{gg}}(k)P_{\textrm{mm}}(k)\left[1-r^{2}(k)\right]} \,,
\end{equation}
where $\textrm{Cov}(K) = N(K)$ is a constant under the assumption of the squeezed limit. In Eq.~(\ref{eq:3d_covariance}), $P_{\textrm{gg}}(k)$ stands for the total galaxy power spectrum including a contribution from shot noise
\begin{equation}
\label{eq:galaxy_power}
	P_{\textrm{gg}}(k) = b^{2}P_{\textrm{cc}}(k)+\frac{1}{\bar{n}_{\textrm{g}}} \,,
\end{equation}
where $P_{\textrm{cc}}(k)$ is the CDM power spectrum and $\bar{n}_{\textrm{g}}$ is the total galaxy number density. Also, to leading order in $f_{\textrm{h}}$, $P_{\textrm{mm}}(k) = P_{\textrm{cc}}(k) + P_{\epsilon_{\textrm{m}}}$ is the total matter power spectrum, \footnote{For simplicity we consider a white noise power spectrum for the matter field, $P_{\epsilon_{\textrm{m}}}$,  which can be absorbed into a redefinition of the galaxy shot noise as in Eq.~(\ref{eq:eff_shot_noise_def}). A generalization to arbitrary (scale-dependent) matter noise is, however, straightforward.} and
\begin{equation}
\label{eq:3d_correlation}
 r(k) = \sqrt{\frac{P_{\textrm{mg}}^{2}(k)}{P_{\textrm{gg}}(k)P_{\textrm{mm}}(k)}} = \left[1+\frac{1}{b^2 \bar{n}_{\textrm{eff}}P_{\textrm{cc}}(k)}\right]^{-1} \,,
\end{equation} 
is the correlation coefficient between matter and galaxy fields, with $P_{\textrm{mg}}(k)$ the matter-galaxy cross power spectrum. The effective number density $\bar{n}_{\textrm{eff}}$ absorbs the matter white noise into the galaxy shot noise as follows
\begin{equation}
\label{eq:eff_shot_noise}
	P_{\textrm{gg}}(k)P_{\textrm{mm}}(k) = \left(b^2 P_{\textrm{cc}}(k) + \frac{1}{\bar{n}_{\textrm{g}}}\right) \left(P_{\textrm{cc}}(k) + P_{\epsilon_{\textrm{m}}}\right) \approx b^2 P_{\textrm{cc}}^{2}(k) + \frac{P_{\textrm{cc}}(k)}{\bar{n}_{\textrm{eff}}} \,,
\end{equation}
where,
\begin{equation}
\label{eq:eff_shot_noise_def}
	\bar{n}_{\textrm{eff}} = \frac{\bar{n}_{\textrm{g}}}{1+ \bar{n}_{\textrm{g}}b^2 P_{\epsilon_{\textrm{m}}}} \,.
\end{equation}
Note that
\begin{equation}
\label{eq:cosmic_variance_free}
	P_{\textrm{gg}}(k)P_{\textrm{mm}}(k)\left[1-r^{2}(k)\right] = \frac{P_{\textrm{cc}}(k)}{\bar{n}_{\textrm{eff}}} \,,
\end{equation}
follows from Eqs.~(\ref{eq:3d_correlation}) and (\ref{eq:eff_shot_noise}). With these ingredients we can write down for the cumulative signal-to-noise
\begin{equation}
\label{eq:snr_3d}
	\textrm{SNR}_{\textrm{3d}}^2 = V \int \frac{d^{3}\vec{K}}{(2\pi)^{3}} \frac{P_{\phi_{\textrm{rel}}\phi_{\textrm{rel}}}(k)}{N(k)} = \int d\log K \  \textrm{SNR}_{\textrm{3d}}^{2}\Big|_{K} \,,  \,
\end{equation}
where
\begin{equation}
\label{eq:snrint_3d}
	\textrm{SNR}_{\textrm{3d}}^{2}\Big|_{K} \equiv V \frac{K^{3}}{2\pi^{2}} \frac{P_{\phi_{\textrm{rel}}\phi_{\textrm{rel}}}(K)}{N(K)} \,.
\end{equation}

In Sec.~\ref{sec:wakes} we explained why the relative velocity reconstruction from HDM wakes is cosmic variance free. This is manifested in Eq.~(\ref{eq:3d_correlation}) by the fact that deviations from a unity cross correlation coefficient, of $r=1$, can be made arbitrarily small in the limit of infinite number density of tracers, $\bar{n}_{\textrm{eff}} \to \infty$. From Eqs.~(\ref{eq:3d_covariance}) and (\ref{eq:snr_3d}) it follows that $\textrm{SNR}^{2} \sim (1-r^2)^{-1}$, which is the expected scaling from cosmic variance cancellation \cite{McDonald:2008sh, Seljak:2008xr}. Note that $\textrm{SNR}^{2} \to \infty$ as $r\to 1$, with $\textrm{SNR}^{2} \propto \bar{n}_{\textrm{eff}}$ as can be seen from Eq.~(\ref{eq:cosmic_variance_free}). Eq.~(\ref{eq:snr_3d}) also has a very simple scaling with the fractional contribution of hot to total matter density, $\textrm{SNR}^2 \propto f_{\textrm{h}}^2$ (under the assumption of $f_{\textrm{h}} \ll 1$ \footnote{The requirement of $f_{\textrm{h}} \ll 1$ was imposed in Sec.~\ref{sec:wakes} to ensure that gravitational backreactions from HDM can be neglected, for simplicity of theoretical modeling. Backreaction effects are expected to emerge when $f_{\textrm{h}} \sim 1$, and can be studied with simulations. It is clear that wakes effects should disappear in the limit of $f_{\textrm{h}} \to 1$, as there are no relative bulk flows in this case. From the observational point of view, fractional contributions of HDM (with a free-streaming scale of cosmological relevance) are constrained to be small due to large-scale structure probes \cite{Peters:2023asu, Garcia-Gallego:2025kiw, Euclid:2024pwi, Dror:2024ibf, Xu:2021rwg, Dayal:2023nwi, Verdiani:2025jcf, Boyarsky:2008xj}.}). In Eq.~(\ref{eq:snr_3d}), V denotes the survey effective volume.

\begin{figure}
	\centering
	\includegraphics[width=0.75\textwidth]{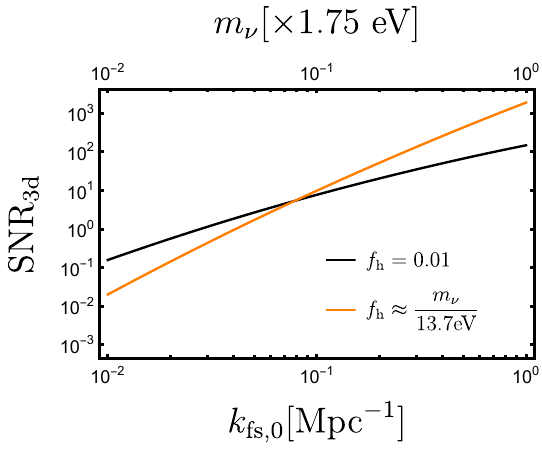}
	\caption{Cumulative signal-to-noise ratio of neutrino and HDM wakes, estimated by the reconstructed relative velocity field from measurements of galaxy-matter cross correlations, as a function of the present-day value of the effective free-streaming scale $k_{\textrm{fs},0}$, assuming a billion spectroscopic redshifts are collected out to $z_{\textrm{max}}=2$ (but see footnote \ref{foot_eff}). The black curve follows from fixing the fractional contribution of hot to total matter density of $f_{\textrm{h}}= 0.01$, and the orange curve corresponds to the case of a single massive neutrino species. These results assume a 3D galaxy survey in combination with a 3D map of the total matter field (or any tracer of HDM, more broadly), at an effective redshift of $z_{\textrm{eff}}=1$.}
	\label{fig:3d_scale}
\end{figure}

In Fig.~\ref{fig:3d_scale} we plot the signal-to-noise ratio as a function of the present-day value of the HDM effective free-streaming scale. We assume an all sky survey collecting redshifts out to $z_{\textrm{max}}=2$, with a corresponding volume of $V= (4\pi/3) x_{\parallel}(z_{\textrm{max}})^{3} \approx 200h^{-3} \textrm{Gpc}^3$ [where $x_{\parallel}(z_{\textrm{max}})$ is the comoving distance to redshift $z_{\textrm{max}}$], and an effective number density of $\bar{n}_{\textrm{eff}} = 5 \times 10^{-3} h^{3} \textrm{Mpc}^{-3}$ \footnote{The effective number density absorbs contributions from the matter noise, i.e. Eq.~(\ref{eq:eff_shot_noise_def}). For example, in the case of weak lensing of galaxies,  $P_{\epsilon_{\textrm{m}}} = \sigma_{\epsilon}^2/2\bar{n}_{\textrm{s}}$ [see Eq.~(\ref{eq:matterps})], with $\bar{n}_{\textrm{s}}$ the number density of source galaxies. It then follows that $\bar{n}_{\textrm{g}}b^2 P_{\epsilon_{\textrm{m}}} = \sigma_{\epsilon}^2/2 \approx 0.07$, where we assume $b=1$, $\bar{n}_{\textrm{g}} = \bar{n}_{\textrm{s}}$ and $\sigma_{\epsilon}=0.37$. We then need a slightly larger galaxy number density of $\bar{n}_{\textrm{eff}} \approx 5.4 \times 10^{-3} h^{3} \textrm{Mpc}^{-3}$ in order to compensate for the matter noise, at fixed $\bar{n}_{\textrm{eff}} = 5 \times 10^{-3} h^{3} \textrm{Mpc}^{-3}$. \label{foot_eff}}. We take all the galaxies to sit at the effective redshift $z_{\textrm{eff}}=1$, for simplicity. The upper x-axis label shows the neutrino mass value associated to a given free-streaming scale, conditioned on Eq.~(\ref{eq:free_streaming}). The black curve assumes a fixed $f_{\textrm{h}}=0.01$, and the orange curve corresponds to a single massive neutrino species satisfying Eq.~(\ref{eq:mnu_abundance}). 

\begin{figure}
	\centering
	\includegraphics[width=0.75\textwidth]{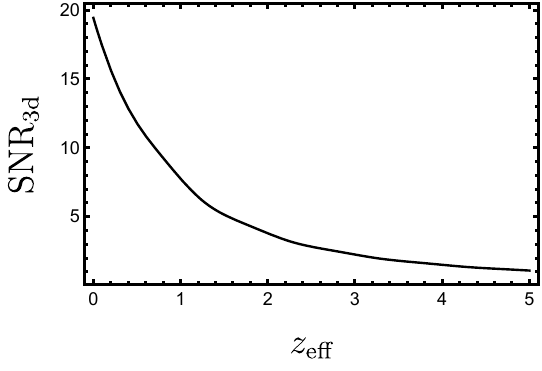}
	\caption{Cumulative signal-to-noise ratio of HDM wakes, estimated by the reconstructed relative velocity field from measurements of galaxy-matter cross correlations, as a function of the effective redshift under the same set-up as Fig.~\ref{fig:3d_scale}, for an effective free-streaming scale $k_{\textrm{fs},0}=0.1\textrm{Mpc}^{-1}$ and galaxy bias $b(z_{\textrm{eff}})=1+z_{\textrm{eff}}$.}
	\label{fig:3d_redshift}
\end{figure}

\begin{figure}
	\centering
	\includegraphics[width=0.75\textwidth]{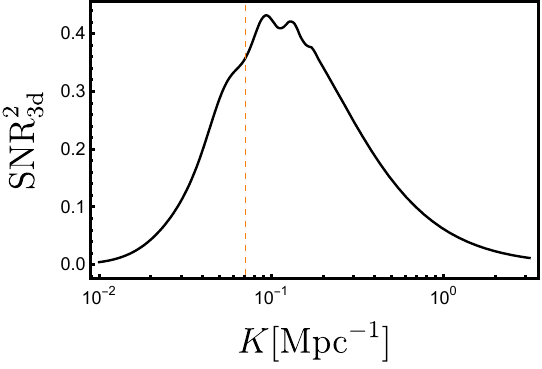}
	\caption{Signal-to-noise ratio of HDM wakes as a function of the reconstructed long-wavelength relative velocity mode [as defined in Eq.~(\ref{eq:snrint_3d})], measured from galaxy-matter cross correlations, and normalized to the cumulative signal-to-noise. The vertical dotted (orange) line shows the effective free-streaming scale at the effective redshift of the galaxy sample, and the set-up is the same as in Fig.~\ref{fig:3d_redshift}.}
	\label{fig:scale_dependence}
\end{figure}

In this set-up HDM wakes become observable when $k_{\textrm{fs},0} \gtrsim 0.03\textrm{Mpc}^{-1}$, or $m_{\nu} \gtrsim 0.04\textrm{eV}$ in the neutrino mass case. This assumes a 3D map of the total matter field (or a tracer of HDM more broadly) with a linear galaxy bias $b(z_{\textrm{eff}})=1+z_{\textrm{eff}}$. The signal-to-noise shows a steep decrease with increasing $z_{\textrm{eff}}$, as illustrated in Fig.~\ref{fig:3d_redshift}, which assumes $k_{\textrm{fs},0}=0.1\textrm{Mpc}^{-1}$. In Fig.~\ref{fig:scale_dependence} we also show the signal-to-noise squared as a function of the long-wavelength mode $K$, as defined in Eq.~(\ref{eq:snrint_3d}). It peaks around the relative velocity coherence scale, which roughly traces the effective free-streaming scale, $k_{\textrm{fs}}(z_{\textrm{eff}}) = k_{\textrm{fs},0}/\sqrt{1+z_{\textrm{eff}}} \approx 0.07\textrm{Mpc}^{-1}$, shown as the vertical dotted (orange) line (after taking $z_{\textrm{eff}}=1$ and $k_{\textrm{fs},0}=0.1\textrm{Mpc}^{-1}$).

Also note from Fig.~\ref{fig:scale_dependence} that there are significant contributions to the signal-to-noise from large values of $K \sim k_{\textrm{fs}}$, which highlights the need for a more complete characterization of the bispectrum away from the squeezed limit. This can be achieved, for example, with measurements of the full bispectrum from N-body simulations. An alternative route could be to explore the mildly non-linear regime of structure formation with perturbative methods, based on the EFTofLSS \cite{Carrasco:2012cv, Baumann:2010tm}, which was recently extended to mixed dark matter scenarios \cite{Verdiani:2025jcf}. However, these existing methods model the dynamics of hot dark matter wakes as a fluid, and hence may not properly capture wakes effects. One can, in principle, develop an EFT approach to mixed dark matter scenarios based on the full Boltzmann equation (e.g. along the lines of \cite{Senatore:2017hyk}). In this case the effects of HDM wakes should be captured by the formalism. We leave these lines of investigation for future work. 

Based on the assumed survey volume and number density, the total effective number (see footnote \ref{foot_eff}) of galaxies would be $N_{\textrm{eff}}=V\bar{n}_{\textrm{eff}} \approx 10^{9}$. From the survey specifications detailed in Sec.~\ref{sec:2d}, we can already expect the Vera Rubin Observatory LSST to collect photometric redshifts of $N_{\textrm{g}} \approx 2.3 \times 10^{9}$ galaxies out to $z=1$. It then seems plausible that a futuristic spectroscopic survey will be able to collect redshifts of about a billion galaxies in the nearby universe \cite{Dodelson:2016wal}. One can of course consider using a sample of galaxies with photometric redshifts due to their typically larger sample size. However, the resulting redshift errors impose significant penalties in signal-to-noise. Since the signal-to-noise mostly comes from scales around $k_{\textrm{fs}}$, photo-z errors become a nuisance whenever $k_{\textrm{fs}} \gtrsim H/\sigma_{z} \approx 0.007 \textrm{Mpc}^{-1}$, which holds for almost all observable scales. We assume LSST-like photometric redshift errors $\sigma_{z}/1+z = 0.03$ \cite{LSSTDarkEnergyScience:2018jkl}, and take $z_{\textrm{eff}}=1$. In our fiducial example of $k_{\textrm{fs,0}}=0.1\textrm{Mpc}^{-1}$, we find that photometric redshift errors decrease the cumulative signal-to-noise from $\textrm{SNR} \approx 7.8$ to $\textrm{SNR} \approx 0.52$, i.e. by a factor of $\sim 15$ \footnote{We model the effect of photo-z errors by a Gaussian damping of the galaxy field modes along the light of sight, $\delta_{\textrm{g}}(\vec{k}) \to e^{-\frac{\sigma_{z}^{2}}{2H^{2}} k_{\parallel}^{2}} \delta_{\textrm{g}}(\vec{k})$. We also apply a similar damping to the relative velocity potential field, since we assume these large scale modes to be traced by a sample of galaxies, $\phi_{\textrm{rel}}(\vec{K}) \to e^{-\frac{\sigma_{z}^{2}}{2H^{2}} K_{\parallel}^{2}} \phi_{\textrm{rel}}(\vec{K})$. Note that the integrals in Eqs.~(\ref{eq:3d_covariance}) and (\ref{eq:snr_3d}) need to be modified, allowing for a splitting of Fourier modes into their components parallel and perpendicular to the line-of-sight.}.

While Fig.~\ref{fig:3d_scale} reveals that HDM wakes are in principle detectable with futuristic experiments at a high signal-to-noise, in the case of $k_{\textrm{fs},0} \gtrsim 0.03\textrm{Mpc}^{-1}$ (assuming $f_{\textrm{h}}=0.01$), this is idealized and may not be achievable in practice. This is because there are no good candidates for a direct 3D tracer of HDM on cosmological scales with near future experiments, and we expect to find significant penalties (in terms of signal-to-noise loss) when considering a 2D tracer such as weak gravitational lensing, which is sensitive to the total matter field and hence traces HDM as well as CDM. The reason for this is as follows. We showed in Sec.~\ref{sec:wakes} that relative velocity reconstruction from the measurement of HDM wakes is cosmic variance free, since $\vec{v}_{\textrm{rel}} \sim \hat{P}^{\textrm{dip}}_{\textrm{ch}}(\vec{k})/\hat{P}^{\textrm{mon}}_{\textrm{ch}}(k)$ is insensitive to the specific realization of the primordial fluctuations. However, in a 2D survey we do not have direct access to a given Fourier mode, only to projections along the line of sight which mixes different modes together. As a consequence, the cosmic variance cancellation will only be partially achieved, and the signal-to-noise ratio reduced. However, it is not apriori clear whether or not HDM will be observable for larger values of the effective free-streaming scale, so we now move on to address this question quantitatively.  

\subsection{Detectability with 2D maps of HDM}
\label{sec:2d}

We now consider combining a 2D galaxy survey with a realistic 2D map of a HDM tracer, which we assume is produced from galaxy lensing for concreteness \cite{Kaiser:1991qi, Bartelmann:2016dvf}. Our observables are now defined on the sphere via projections along the line-of-sight
\begin{equation}
\label{eq:2d_obs}
	A(\vec{n}) = \int dx_{\parallel} W_{A}(x_{\parallel}) A(x_{\parallel}, \vec{x}) \,,
\end{equation} 
where $\vec{x} = \vec{x}_{\parallel} + \vec{x}_{\perp} = x_{\parallel}(\hat{z}+\vec{n})$, with $\hat{z}$ the line-of-sight direction and $x_{\parallel}$ (or $\vec{x}_{\perp}$) denote comoving distances parallel (or perpendicular) to the line-of-sight. In Eq.~(\ref{eq:2d_obs}), $A=\{\kappa, \delta_{g}, \phi_{\textrm{rel}}\}$ with $\kappa(x_{\parallel}, \vec{x})$ the lensing convergence, $\delta_{\textrm{g}}(x_{\parallel}, \vec{x})$ the galaxy density contrast and $\phi_{\textrm{rel}}(x_{\parallel}, \vec{x})$ the relative velocity scalar potential. The lensing kernel is given by
\begin{equation}
\label{eq:lensing_kernel}
	W_{\kappa}(x_{\parallel}) = \frac{3}{2} \Omega_{\textrm{m}} H_{0}^{2} \frac{x_{\parallel}}{a(x_{\parallel})} \int_{x_{\parallel}}^{\infty} dx_{\parallel}' \frac{n_{s}(x_{\parallel})}{\bar{n}_{s}} \frac{x_{\parallel}'-x_{\parallel}}{x_{\parallel}'} \,,
\end{equation}
and we will specify the kernels $W_{\textrm{g}}(x_{\parallel})$ and $W_{\phi_{\textrm{rel}}}(x_{\parallel})$ shortly \footnote{For simplicity of notation we set $W_{\delta_{\textrm{g}}} \equiv W_{\textrm{g}}$.}. In Eq.~(\ref{eq:lensing_kernel}), $n_{s}(x_{\parallel})$ stands for the (angular) number density of source galaxies in the cosmic shear survey, while $\bar{n}_{s} = \int dx_{\parallel} n_{s}(x_{\parallel})$ is the total number of source galaxies.

In our forecasts we adopt survey specifications from the Vera Rubin Observatory LSST , i.e., $n_{s}(z) \propto 1/(2z_{0}) (z/z_{0})^{2} e^{-z/z_{0}}$ with $z_{0}=0.3$ and $\bar{n}_{s} = 40 \  \textrm{arcmin}^{-2}$ \cite{abell2009lsst, Yu:2018tem} (this distribution is plotted in Fig.~\ref{fig:2d_redshift} as the orange dashed curve, out to $z=1.5$ and normalizing its peak value to unity). We assume a survey area of $18,000 \ \textrm{deg}^{2}$, corresponding to $f_{\textrm{sky}} \approx 0.4$. The angular number density of $\bar{n}_{s} = 40 \  \textrm{arcmin}^{-2}$ than translates to a total  number of $N_{\textrm{g}} \approx 2.4 \times 10^{9}$ galaxies. 

We also adopt the same distribution for galaxy clustering, but break it down into a varying number $N_{\textrm{z-bins}}$ of equally spaced (non-overlapping) redshift bins in the interval $0 \leq z \leq 1.5$. We will take the case of $N_{\textrm{z-bins}}=6$ as a baseline, for which each redshift bin has a width of $\Delta z=0.25$ \footnote{We do not include galaxies at $z>1.5$ since the signal-to-noise ratio decays rather steeply with redshift as shown in Figs.~\ref{fig:3d_redshift} and \ref{fig:2d_redshift}. As an example, in the case of $k_{\textrm{fs},0}=0.1 \textrm{Mpc}^{-1}$ and $f_{\textrm{h}} = 0.01$, we find that going out to a $z_{\textrm{max}}=2$ with $N_{z-\textrm{bins}} = 8$ (which adds two more redshift bins of the same width to the fiducial scenario of $z_{\textrm{max}}=1.5$ and $N_{z-\textrm{bins}} = 6$) only increases the signal-to-noise by about $\approx 0.2\%$.}. The total galaxy number density in the i-th redshift bin is given by 
\begin{equation}
\label{eq:total_galaxies_bin}
	\bar{n}_{\textrm{g},i} = \bar{n}_{s} \int_{z_{\textrm{min},i}}^{z_{\textrm{max},i}} dz \frac{1}{2z_{0}} \left(\frac{z}{z_{0}}\right)^{2} e^{-\frac{z}{z_{0}}} \,,
\end{equation}
where $z_{\textrm{min},i}$ ($z_{\textrm{max},i}$) defines the lower (upper) limit of the i-th bin. Our approach will follow \cite{Sherwin:2015baa, Schmittfull:2017ffw}, optimally combining the different redshift bins into a single galaxy and relative velocity potential two-dimensional fields while maximizing the total signal-to-noise (this procedure is detailed in Appendix \ref{sec:app3}). This can be achieved by letting $W_{\textrm{g}} \to \sum_{i} c_{i}W_{\textrm{g},i}$ and $W_{\phi_{\textrm{rel}}} \to \sum_{i} d_{i}W_{\textrm{g},i}$, where 
\begin{equation}
 \label{eq:galaxy_kernel}
	 W_{\textrm{g},i}(x_{\parallel}) = b(x_{\parallel}) \frac{n_{\textrm{g},i}(x_{\parallel})}{\bar{n}_{\textrm{g},i}} \frac{dz}{dx_{\parallel}} \,.
\end{equation}
In Eq.~(\ref{eq:galaxy_kernel}), $n_{\textrm{g},i}$ stands for the redshift distribution in the i-th redshift bin, which we assume is identical to $n_{s}$ but with a nonzero support in the interval $z_{\textrm{min},i} \leq z \leq z_{\textrm{max},i}$.

We next perform a multipole decomposition of Eq.~(\ref{eq:2d_obs}). In the flat sky approximation this is given by the 2D Fourier transform
\begin{equation}
\label{eq:2d_fourier}
	A(\vec{\ell}) = \int d^{2}\vec{n}  \ e^{-i\vec{n} \cdot \vec{\ell}} A(\vec{n}) = \int \frac{dx_{\parallel}}{x_{\parallel}^{2}}  W_{A}(x_{\parallel}) \int \frac{dk_{\parallel}}{2\pi} e^{ix_{\parallel} k_{\parallel}} A\left(x_{\parallel},\vec{k}_{\parallel} + \frac{\vec{\ell}}{x_{\parallel}}\right) \,.
\end{equation}

It is now a straightforward exercise to compute the two-dimensional analog of the bispectrum in Eq.~(\ref{eq:bispectrum}). Using Eqs.~(\ref{eq:bispectrum_calculated}) and (\ref{eq:2d_fourier}), we obtain under the Limber approximation \cite{Limber:1954zz, LoVerde:2008re}
\begin{equation}
\label{eq:2d_bispectrum}
	\frac{1}{2} \left\langle \left[\delta_{\textrm{g}}(\vec{\ell})\kappa(\vec{\ell '}) - \kappa(\vec{\ell})\delta_{\textrm{g}}(\vec{\ell '})\right] \phi_{\textrm{rel}}(-\vec{L}) \right\rangle = (2\pi)^{2} \delta^{2}(\vec{\ell}+\vec{\ell '} - \vec{L}) B(\vec{\ell},\vec{L}) \,,
\end{equation}
where,
\begin{equation}
\label{eq:2d_bispectrum_2}
\small
	B(\vec{\ell},\vec{L}) = f_{\textrm{h}} (\hat{\ell} \cdot \hat{L}) \int \frac{dx_{\parallel}}{x_{\parallel}^{4}} W_{\textrm{g}}(x_{\parallel}) W_{\kappa}(x_{\parallel}) W_{\phi_{\textrm{rel}}}(x_{\parallel}) \frac{\gamma}{\sigma_{\nu}(x_{\parallel})} T_{\textrm{dip}}\left(x_{\parallel}, \frac{\ell}{x_{\parallel}}\right) P_{\textrm{ch}}\left(x_{\parallel}, \frac{\ell}{x_{\parallel}}\right) P_{\phi_{\textrm{rel}}\phi_{\textrm{rel}}}\left(x_{\parallel}, \frac{L}{x_{\parallel}}\right) \,.
\end{equation}
We can change the integration variable from the distance along the line-of-sight to redshift, using $dx_{\parallel} = dz/H(z)$ and  $W_{A}(x_{\parallel}) = (dz/dx_{\parallel}) W_{A}(z)$. We can also write the following formula for 2D (angular) power spectra under the Limber approximation
\begin{equation}
\begin{split}
\label{eq:2d_ps}
	& P_{\textrm{ch}}(l) = \int dz \frac{H(z)}{x_{\parallel}^{2}(z)} W_{g}(z) W_{\kappa}(z) P_{\textrm{ch}} \left(z, \frac{\ell}{x_{\parallel}(z)}\right) \\ & P_{\phi_{\textrm{rel}}\phi_{\textrm{rel}}}(L) = \int dz \frac{H(z)}{x_{\parallel}^{2}(z)} \left[W_{\phi_{\textrm{rel}}}(z)\right]^{2} P_{\phi_{\textrm{rel}}\phi_{\textrm{rel}}} \left(z, \frac{L}{x_{\parallel}(z)}\right) \,.
\end{split}
\end{equation}
Combining these various ingredients, we may now rewrite Eq.~(\ref{eq:2d_bispectrum_2}) in the form
\begin{equation}
\label{eq:2d_bispectrum_3}
	B(\vec{\ell},\vec{L}) = f_{\textrm{h}} \alpha_{\ell, L} (\hat{\ell} \cdot \hat{L}) P_{\textrm{ch}}(\ell) P_{\phi_{\textrm{rel}}\phi_{\textrm{rel}}}(L) \,,
\end{equation}
where
\begin{equation}
\label{eq:2d_coefficient}
\small
	\alpha_{\ell, L} = \frac{1}{P_{\textrm{ch}}(l)P_{\phi_{\textrm{rel}}\phi_{\textrm{rel}}}(L)} \int dz \frac{H^{2}(z)}{x_{\parallel}^{4}(z)} W_{\textrm{g}}(z) W_{\kappa}(z) W_{\phi_{\textrm{rel}}}(z) \frac{\gamma}{\sigma_{\nu}(z)} T_{\textrm{dip}}\left(z, \frac{\ell}{x_{\parallel}(z)}\right) P_{\textrm{ch}}\left(z, \frac{\ell}{x_{\parallel}(z)}\right) P_{\phi_{\textrm{rel}}\phi_{\textrm{rel}}}\left(z, \frac{L}{x_{\parallel}(z)}\right) \,,
\end{equation}
is a dimensionless coefficient that accounts for redshift dispersion in a given bin. In order to write down a quadratic estimator (in analogy to the 3D case discussed in Sec.~\ref{sec:3d}) we introduce the 2D analog of Eq.~(\ref{eq:pre_squeezed_bispectrum}), which captures the dipole distortion to the local cross correlation in the presence of a HDM-CDM relative velocity perpendicular to the line-of-sight, with scalar potential $\phi_{\textrm{rel}}(\vec{L})$,
\begin{equation}
\label{eq:dipole_dis_2d}
	\langle \delta_{\textrm{g}}(\vec{\ell}) \kappa(\vec{\ell '}) \rangle_{\textrm{dip}} = f_{\textrm{h}} \alpha_{\ell,L} (\hat{\ell} \cdot \hat{L}) \phi_{\textrm{rel}}(\vec{L}) P_{\textrm{ch}}(\ell) \,.
\end{equation}

We are now ready to write the estimator for $\phi_{\textrm{rel}}(\vec{L})$. The 2D analog of Eq.~(\ref{eq:3d_estimator}) is
\begin{equation}
\label{eq:2d_estimator}
	\hat{\phi}_{\textrm{rel}}(\vec{L}) = N(L) \int \frac{d^2 \vec{\ell}}{(2\pi)^{2}} W_{A}(\vec{\ell},\vec{\ell ' }) \kappa(\vec{\ell}) \delta_{\textrm{g}}(\vec{\ell '}) \,.
\end{equation}
Once again the filter function $W_{A}(\vec{\ell},\vec{\ell ' })$ and covariance $N(L)$ are to be determined from the requirement that Eq.~(\ref{eq:2d_estimator}) is an unbiased optimal (minimum noise) estimator. The final formula for the cumulative signal-to-noise ratio, derived in Appendix \ref{sec:app2}, reads
\begin{equation}
\label{eq:2d_stn}
\textrm{SNR}^{2}_{\textrm{2d}} = \frac{1}{2\pi} f_{\textrm{sky}} f_{\textrm{h}}^{2} \int LdL \int \ell d\ell \ \alpha_{\ell,L}^{2} \frac{ \left[P_{\textrm{ch}}(\ell)\right]^{2} P_{\phi_{\textrm{rel}}\phi_{\textrm{rel}}}(L) }{P_{\kappa\kappa}(\ell)P_{\textrm{g} \textrm{g}}(\ell)[1-r^{2}(\ell)]} \,,
\end{equation}
where the total matter angular power spectrum is, to leading order in $f_{\textrm{h}}$,
\begin{equation}
\label{eq:matterps}
	P_{\kappa\kappa}(\ell) = \int dz \frac{H(z)}{x_{\parallel}^{2}(z)} \left[ W_{\kappa}(z)\right]^2 P_{\textrm{cc}} \left(z, \frac{\ell}{x_{\parallel}(z)}\right) + \frac{\sigma_{\epsilon}^{2}}{2\bar{n}_{s}} \,.
\end{equation}

In Eq.~(\ref{eq:matterps}) we account for noise in the matter field via the intrinsic ellipticity dispersion of weak lensing sources, which we take to be $\sigma_{\epsilon}=0.37$ \cite{Euclid:2023ove, DES:2020ypx}. The total galaxy angular power spectrum is
\begin{equation}
	\label{eq:galaxyps}
	P_{\textrm{gg}}(\ell) = \int dz \frac{H(z)}{x_{\parallel}^{2}(z)} \left[ W_{\textrm{g}}(z)\right]^2 P_{\textrm{cc}} \left(z, \frac{\ell}{x_{\parallel}(z)}\right) + \frac{1}{\bar{n}_{\textrm{g}}} \,,
\end{equation}
including shot noise. The matter-galaxy cross spectrum is
\begin{equation}
\label{eq:galaxyxmatter}
	P_{\textrm{gm}}(\ell) = \int dz \frac{H(z)}{x_{\parallel}^{2}(z)}  W_{\textrm{g}}(z) W_{\kappa}(z)  P_{\textrm{cc}} \left(z, \frac{\ell}{x_{\parallel}(z)}\right) \,.
\end{equation}
Finally, the galaxy-galaxy lensing correlation coefficient $r(l)$ appearing in Eq.~(\ref{eq:2d_stn}) is
\begin{equation}
\label{eq:2d_corr_coefficient}
	r(l) = \sqrt{\frac{	P_{\textrm{g}\kappa}^{2}(\ell)}{P_{\textrm{gg}}(\ell) P_{\kappa\kappa}(\ell)}} \,.
\end{equation}

Note that $r(l)$ does not approach unity as $\bar{n}_{s},\bar{n}_{\textrm{g}} \to \infty$, so that the relative velocity reconstruction is now limited by sample variance, in agreement with our expectation from the discussion at the end of Sec.~\ref{sec:3d}. This is due to the presence of nonzero stochasticity between the lensing convergence and galaxy fields. 

\begin{figure}
	\centering
	\includegraphics[width=0.75\textwidth]{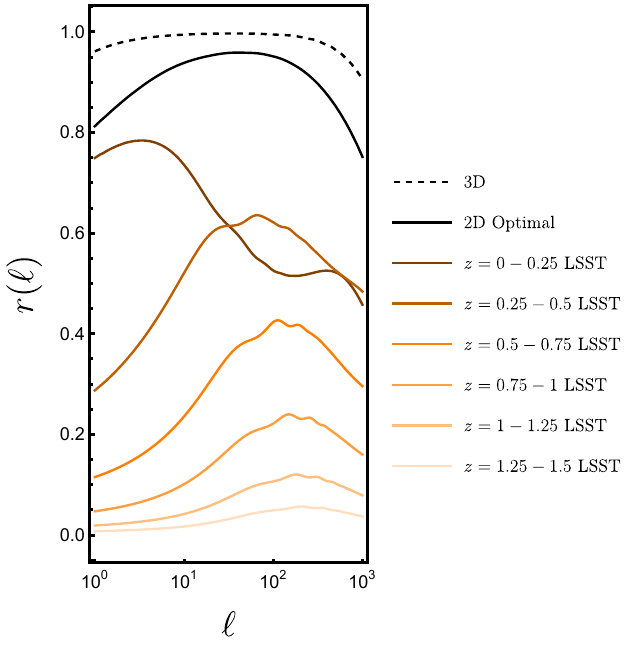}
	\caption{Correlation coefficient between galaxy and lensing converge fields. The lines on different shades of orange correspond to varying redshift bins (for the galaxy sample) in the 2D case with a total number of $N_{\textrm{z-bins}}=6$ equally spaced bins in the interval $0\leq z \leq 1.5$.  The solid black line corresponds to the optimal combination of redshift bins.  In all cases the lensing converge field is computed with the entire (LSST mock) source galaxy population out to $z=5$. We also include for comparison the three-dimensional galaxy-matter correlation coefficient, evaluated at $k=\ell/\chi(z_{\textrm{eff}}=1)$, as the dashed black line.}
	\label{fig:optimal_correlation}
\end{figure}

For our forecasts we combine the different redshift bins into a single galaxy and relative velocity potential fields, while maximizing the correlation coefficient $r(l)$ and the signal-to-noise per long-wavelength mode. We take
\begin{equation}
\label{eq:optimal_1}
	\delta_{\textrm{g}} = \sum_{i=1}^{N_{\textrm{z-bins}}} c^{\ell}_{i} 	\delta_{\textrm{g}_{i}} \,,
\end{equation}
and
\begin{equation}
	\label{eq:optimal_2}
	\phi_{\textrm{rel}} = \sum_{i=1}^{N_{\textrm{z-bins}}} d^{\ell,L}_{i} 	\delta_{\textrm{g}_{i}} \,,
\end{equation}
where the coefficients $c^{\ell}_{i}$ and $d^{\ell,L}_{i}$ are obtained as solutions to an optimization problem described (and solved) in Appendix \ref{sec:app3}. The resulting optimal correlation coefficient is plotted in Fig.~\ref{fig:optimal_correlation}, for our baseline choice of $N_{\textrm{z-bins}}=6$, against the correlation coefficient for the individual redshift bins. We also add the three dimensional correlation coefficient from Eq.~(\ref{eq:3d_correlation}) for comparison, evaluated at $k=\ell/\chi(z_{\textrm{eff}}=1)$ with the survey specifications discussed in Sec.~\ref{sec:3d}. The optimal 2D combination allows for a more effective sample variance cancellation when compared to any of the individual redshift bins, but it is still not close to the levels of cancellation achieved in the three-dimensional case, where the deviations from $r=1$ are solely due to a finite number of tracers and disappear entirely in the limit of $\bar{n}_{\textrm{g}} \to \infty$.

Some representative numerical values for the cumulative signal-to-noise ratio are shown in Tab.~\ref{table:stn_tab}, under the optimal combination of redshift bins for the galaxy sample, for our baseline choice of $N_{\textrm{z-bins}}=6$. The first three rows assume $f_{\textrm{h}}=0.01$, while the last three correspond to the case of a (single) massive neutrino species where $f_{\textrm{h}}$ and $k_{\textrm{fs},0}$ are determined by Eqs.~(\ref{eq:mnu_abundance}) and (\ref{eq:free_streaming}). The first column shows the total signal-to-noise while the second assumes $\bar{n}_{s},\bar{n}_{\textrm{g}} \to \infty$, and hence corresponds to the cosmic variance (CV) limit. As we can see, the total signal-to-noise remains below unity throughout the entire parameter space considered, and hence HDM wakes are unlikely to be ever observed with a 2D tracer of HDM on cosmological scales.

\begin{table}
	\centering
	\begin{tabular}{|c|c|c|c|}
		\hline
		& $\textrm{SNR}$ & $\textrm{SNR} \  (\textrm{CV})$  \\ \hline
		$k_{\textrm{fs},0}=0.01\textrm{Mpc}^{-1}$ & $9.0 \times 10^{-4}$ & $1.2 \times 10^{-3}$  \\ \hline
		$k_{\textrm{fs},0}=0.1\textrm{Mpc}^{-1}$ & $1.8 \times 10^{-2}$ & $3.0 \times 10^{-2}$  \\ \hline
		$k_{\textrm{fs},0}=1\textrm{Mpc}^{-1}$	& $5.7 \times 10^{-2}$ & $1.4 \times 10^{-1} $   \\ \hline
		$m_{\nu} = 0.1\textrm{eV}$ & $7.3 \times 10^{-3} $ & $1.1 \times 10^{-2} $\\ \hline
		$m_{\nu} = 0.5\textrm{eV}$ & $1.4 \times 10^{-1} $ & $2.8 \times 10^{-1} $ \\ \hline
		$m_{\nu} = 1\textrm{eV}$ &  $3.7 \times 10^{-1} $ & $8.5 \times 10^{-1}$ \\ \hline
	\end{tabular}
	\caption{Cumulative signal-to-noise ratios for neutrino and HDM wakes, estimated by the reconstructed relative velocity field perpendicular to the light-of-sight, from measurements of 2D galaxy-matter cross correlations. We optimally combine $N_{\textrm{z-bins}}=6$ equally space redshift bins in the interval $0\leq z \leq 1.5$ for the galaxy sample (the relative signal-to-noise from each individual bin can be found in Fig.~\ref{fig:2d_redshift}). The first three rows assume a fixed fractional contribution of a hot subcomponent of the total dark matter, $f_{\textrm{h}}=0.01$, for varying present-day values of the effective free-streaming scale $k_{\textrm{fs},0}$. The last three rows correspond to the case of a single massive neutrino species. The second column takes the limit of infinite number density of tracers, hence revealing the cosmic variance limit. In all cases $\textrm{SNR} \ll 1$.}
	\label{table:stn_tab}
\end{table} 

The survey specifications considered here, typical of upcoming experiments, are already enough to almost reach the cosmic variance limit (at least for smaller values of the effective free-streaming wavenumber, see Tab.~\ref{table:stn_tab}). This means that detection prospects will not improve with futuristic surveys of even higher sensitivity. This behavior is opposite to the one seen in the 3D case, where  $\textrm{SNR}^{2} \propto \bar{n}_{\textrm{g}}$. We also note that the differences in signal-to-noise between the first and second columns of Tab.~\ref{table:stn_tab} are mostly driven by the noise in the matter field (due to the intrinsic ellipticity of weak lensing sources), rather than shot noise in galaxy clustering \footnote{In other words, sending $\bar{n}_{\textrm{g}} \to \infty$ at fixed $\bar{n}_{s}$ produces essentially the same results.}. 

\begin{figure}
	\centering
	\includegraphics[width=0.75\textwidth]{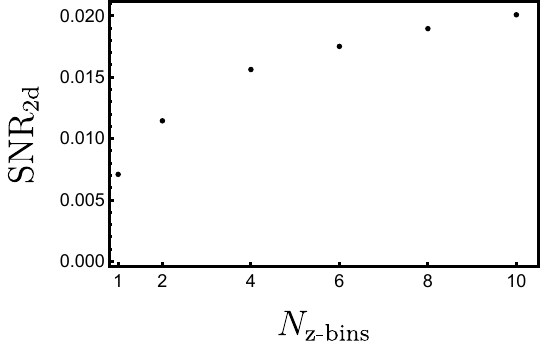}
	\caption{Signal-to-noise ratios for neutrino and HDM wakes as a function of the number of tomographic redshift bins $N_{\textrm{z-bins}}$, as estimated by the reconstructed relative velocity field perpendicular to the light-of-sight, from measurements of 2D galaxy-matter cross correlations. We combine the different bins into single optimal two-dimensional fields [for both galaxy and relative velocity potential, see Eqs.~(\ref{eq:optimal_1}) and (\ref{eq:optimal_2})]. The $N_{\textrm{z-bins}} \to \infty$ limit should reproduce the 3D case, up to some caveats discussed in the main text. We assume a fractional contribution of hot to total matter density of $f_{\textrm{h}}=0.01$, a present-day value for the effective free-streaming scale of $k_{\textrm{fs},0}=0.1\textrm{Mpc}^{-1}$, and an LSST-like galaxy redshift distribution with a total number density of $\bar{n} = 40 \ \textrm{arcmin}^{-2}$ [for the source galaxies used for lensing as well, see lines of text around Eqs.~(\ref{eq:lensing_kernel}) and (\ref{eq:total_galaxies_bin})].}
	\label{fig:tomographic}
\end{figure}

\begin{figure}
	\centering
	\includegraphics[width=0.75\textwidth]{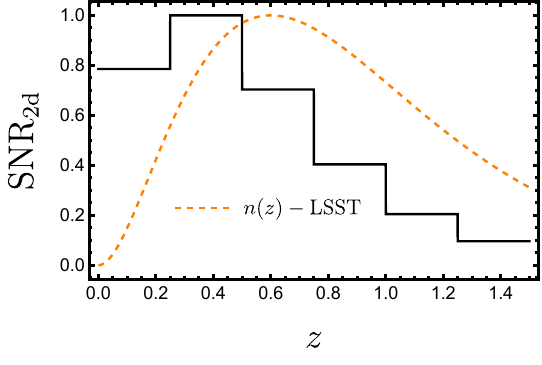}
	\caption{Similar setup as in  Fig.~\ref{fig:tomographic}, now showing signal-to-noise for different redshift bins in black, normalized to a unity peak. The orange dashed curve shows the assumed galaxy redshift distribution, which is also similarly normalized.}
	\label{fig:2d_redshift}
\end{figure}

It is also interesting to examine the dependence of the signal-to-noise on the number of redshift bins $N_{\textrm{z-bins}}$ for the galaxy sample, which illustrates the gains derived from a tomographic approach. This is plotted in Fig.~\ref{fig:tomographic}, for $k_{\textrm{fs},0}=0.1\textrm{Mpc}^{-1}$ and $f_{\textrm{h}}=0.01$. As expected, the signal-to-noise increases with increasing $N_{\textrm{z-bins}}$. This mostly derives from a more effective canceling of sample variance due to a reduced stochasticity between the galaxy and lensing convergence fields when optimally combining a larger number of redshift bins (see Fig.~\ref{fig:optimal_correlation}). 

In principle one could recover the 3D case as the $ N_{\textrm{z-bins}} \to \infty$ limit of the 2D case, but in practice this would be ultimately limited by the finite resolution associated to photometric measurements of galaxy redshifts. Taking $\sigma_{z}/1+z = 0.03$, and approximating $\sigma_{z} \approx  0.06$ (at the effective redshift $z_{\textrm{eff}} \approx 1$), $N_{\textrm{z-bins}} = 1.5/\sigma_{z} \approx 25$ is the threshold above which photo-z errors need to be explicitly accounted for, leading to a dilution of signal-to-noise when compared to our tomographic 2D forecasts \cite{Ma:2005rc}. There are two additional caveats to the statement that the 3D forecasts can be obtained in the $N_{\textrm{z-bins}} \to \infty$ limit. First, our calculations rely on the Limber approximation, and hence completely neglect contributions from Fourier modes along the line-of-sight. As $N_{\textrm{z-bins}}$ increases and the bin width becomes smaller, such contributions increase in relevance and one needs to go beyond the Limber approximation in order to capture them. Second, breaking the galaxy sample into a large number of tomographic bins effectively dilutes the galaxy number density per bin, which we expect should eventually lead to a loss of signal-to-noise. This is in contrast to our 3D forecast which implicitly assumes all galaxies sit at the effective redshift $z_{\textrm{eff}}=1$ \footnote{Of course this is not an intrinsic limitation of the 2D case that is not present in 3D, rather, it just points to our 3D forecast being overly optimistic by not accounting for survey realism.}. 

We also plot the relative signal-to-noise for each individual redshift bin, in the case of $N_{\textrm{z-bins}}=6$, $k_{\textrm{fs},0}=0.1\textrm{Mpc}^{-1}$ and $f_{\textrm{h}}=0.01$. The shape of the histogram shown in black is set by the interplay between the steep redshift dependence seen in Fig.~\ref{fig:3d_redshift} and the galaxy distribution shown as the orange dotted curve. The contributions to the cumulative signal-to-noise from redshift bins above $z \gtrsim 1$ should be marginal.

\subsection{Discussion of results}
\label{sec:discussion}

The general lesson to be learned from Secs.~\ref{sec:3d} and \ref{sec:2d} is that neutrino and HDM wakes are unlikely to be ever observed with 2D maps of the large-scale matter field (see Tab.~\ref{table:stn_tab}), which is the most natural tracer of a hot subcomponent of the total dark matter. On the other hand, HDM wakes are in principle observable with 3D maps of a tracer of HDM, provided that the effective free streaming wavenumber is sufficiently large, i.e. $k_{\textrm{fs},0} \gtrsim 0.03 \textrm{Mpc}^{-1}$ (see Fig.~\ref{fig:3d_scale}).

Many separate ingredients combine to produce the large differences in signal-to-noise between these two cases. The most relevant of them is the more pronounced sample variance cancellation seen in the 3D case, highlighted in Fig.~\ref{fig:optimal_correlation} on the correlation coefficient between the galaxy and lensing convergence. The limit of $r \to 1$ corresponds to no stochasticity between these two fields, and exact sample variance cancellation. On top of this, in the 2D case we can only reconstruct the components of the relative velocity perpendicular to the line-of-sight, i.e. modes along the light-of-sight are lost \footnote{This information can be partially recovered by tomography and going beyond the Limber approximation (which can lead to substantial improvements in some specific scenarios, e.g. \cite{Shen:2025nat}).}.

While neutrino and HDM wakes on the large-scale structure are in principle observable with 3D maps of a tracer of HDM, potential candidates for such a tracer are limited. The typical example of a 3D survey, i.e. spectroscopic galaxy positions, is expected to mostly trace the distribution of CDM and baryons, but not a HDM species which is homogeneously distributed on the typical scales associated to galaxy formation processes \footref{foot_2} (assuming an effective free-streaming scale of cosmological relevance, which we take to be the defining feature of a hot dark subcomponent of the total dark matter). To be more precise, let us consider a sample of galaxies (labelled by $\alpha$) while allowing for a HDM contribution to the bias as follows
\begin{equation}
\label{eq:hdm_bias}
	\delta_{g_{\alpha}} = (1-f_{\textrm{h}}) b_{\textrm{c}}^{\alpha} \delta_{\textrm{c}} + f_{\textrm{h}} b_{\textrm{h}}^{\alpha} \delta_{\textrm{h}} \,. 
\end{equation}
We can cross-correlate it with another such galaxy sample (labelled by $\beta$), and schematically consider the antisymmetric combination 
\begin{equation}
\label{eq:cross_galaxies}
	\delta_{g_{\alpha}} \delta_{g_{\beta}} - \delta_{g_{\beta}}\delta_{g_{\alpha}} \approx f_{\textrm{h}} (b_{\textrm{c}}^{\alpha}b_{\textrm{h}}^{\beta} - b_{\textrm{c}}^{\beta}b_{\textrm{h}}^{\alpha})(\delta_{\textrm{c}}\delta_{\textrm{h}} - \delta_{\textrm{h}}\delta_{\textrm{c}}) \,.
\end{equation}   
Then, as argued in \cite{Celik:2025wkt}, $b_{\textrm{h}}^{\alpha} \approx 0$ in the limit of $k_{\textrm{fs}} \ll k_{\textrm{NL}}$ with $k_{\textrm{NL}}$ the scale of nonlinearities, since then anisotropies in the HDM species are washed out on halo scales, and the HDM does not contribute to the galaxy formation process. In the opposite regime, i.e. $k_{\textrm{fs}} \gg k_{\textrm{NL}}$, $b_{\textrm{h}}^{\alpha} \approx b_{\textrm{c}}^{\alpha}$ since now the HDM looks exactly like CDM on halo scales, and Eq.~(\ref{eq:cross_galaxies}) vanishes. The prospect of using only galaxy samples to probe HDM wakes is then limited to an intermediate regime $k_{\textrm{fs}} \sim k_{\textrm{NL}}$, where $0<b_{\textrm{h}}^{\alpha}<b_{\textrm{c}}^{\alpha}$, and the signal-to-noise is suppressed with respect to our 3D estimate of Sec. \ref{sec:3d} by a factor of $\sim b_{\textrm{h}}^{\alpha}/b_{\textrm{c}}^{\alpha}$. A quantitative measurement of HDM bias from N-body simulations has never been carried out in the literature, and would help to elucidate the feasibility of this approach.  

A similar story holds for the velocity bias, and hence redshift-space distortions, where the quadrupole distortion appearing in the standard Kaiser formula \cite{Kaiser:1987qv} is mostly sourced by the velocity field of CDM and baryons \cite{Verdiani:2025znc, marulli2011effects}, but there should be some room for a small contribution from the subdominant HDM \cite{Celik:2025wkt}. Yet another (more speculative) idea to probe HDM wakes with upcoming surveys is the possibility of performing a three-dimensional analysis of weak lensing surveys \cite{LozanoTorres:2022mci, Mandel:2005xh, Heavens:2003jx, Zieser:2016lmz, Pratten:2016ojf, Kitching:2014lga, CFHTLenS:2014rje}.

We emphasize that our 2D forecasts are conditioned on a specific choice of the galaxy redshift distribution, shown as a dashed orange curve in Fig.~\ref{fig:2d_redshift}. Given the steep redshift dependence of the signal-to-noise, illustrated in Fig.~\ref{fig:3d_redshift}, one can expect to be able to increase it significantly by shifting the galaxy distribution towards lower redshifts. However, this would be intrinsically limited by the number of galaxies in the nearby universe, causing an increase in the noise which necessarily follows the boost in signal.

We will finally discuss an intriguing possibility which was not addressed in the previous sections. As shown in Fig.~\ref{fig:3d_scale}, the signal-to-noise increases with the effective free-streaming scale $k_{\textrm{fs},0}$. Additionally, the signal-to-noise peaks at around $k_{\textrm{fs}}(z_{\textrm{eff}})$ as illustrated in Fig.~\ref{fig:scale_dependence}. These two observations suggest the tantalizing possibility of searching for Warm Dark Matter (WDM) wakes in substructures on small scales \cite{Xiao:2024qay, Mondino:2020rkn, Wagner-Carena:2024axc, Mao:2024pic, Nadler:2024ims}, instead of HDM wakes in the large-scale structure. In the neutrino mass case such a search is not very well motivated since an upper bound to individual neutrino masses, i.e. $m_{\nu} \lesssim 1\textrm{eV}$, can be placed very robustly from the requirement that neutrinos are still relativistic at the time of recombination. However, a small-scale search of WDM wakes, more broadly, is very well motivated since many WDM models are compatible with an effective free-streaming scale of astrophysical size, and in this region of parameter space much larger values of $f_{\textrm{h}}$ are compatible with observations (although of course WDM disappear in the limit of $f_{\textrm{h}} \to 1$, since then there are no relative bulk flows). 

That being said, a small-scale search for WDM wakes would also involve many new difficulties. One of them is the limited statistics of objects in the astrophysical environment within and surrounding our own galaxy. Another issue is associated to the challenge of accurately modeling the dynamics on small scales, which require sophisticated theoretical and numerical tools to account for the intricacies of  gravitational nonlinearities and relevant astrophysical processes \footnote{The covariance matrix is also harder to model on small scales due to the increased relevance of non-Gaussian contributions.}. To be concrete within the framework adopted in Sec.~\ref{sec:wakes}, pushing beyond $k_{\textrm{fs},0} \gtrsim 1\textrm{Mpc}^{-1}$ (or even $k_{\textrm{fs},0} \gtrsim 0.1 \textrm{Mpc}^{-1}$ to be more conservative) requires running high resolution N-body simulations including the effects of a warm subcomponent of the dark matter \cite{Sharma:2025ldt, Leo:2017zff, Brandbyge:2017tdc, Hu:2025lrl}, and non-perturbative models for the galaxy-halo connection \cite{Berlind:2001xk}. However, we stress that the relative velocity can be reconstructed from measurements of WDM wakes without having an accurate model for the nonlinear gravitational evolution on small scales (which is informed by the measurement itself) as we explained in Sec.~\ref{sec:wakes} below Eq.~(\ref{eq:nu_velocity}) \footnote{Though one still needs to model the galaxy-halo connection, assuming galaxies are used to trace CDM.}. Still, having an accurate model plays a key role in forecasting the observability of WDM wakes on small-scale structures, and interpreting results from experiments in terms of what they mean for fundamental physics. A careful viability assessment of a small-scale search for WDM wakes is beyond the scope of this work, and is left for future investigation. 

\section{Conclusion}
\label{sec:concl}

Neutrinos preferentially accumulate downstream of moving CDM structures, sourcing neutrino wakes in the large-scale distribution of matter that may be observable with future cosmological surveys. In this work we revisit this effect, now rephrasing it in terms of a broader class of Hot Dark Matter (HDM) models parametrized by the fractional contribution of hot to the total matter density $f_{\textrm{h}}$, and the present-day value of the effective free-streaming scale $k_{\textrm{fs},0}$. We also improve the theoretical modeling of HDM wakes, and discuss detection prospects under more realistic conditions than previously considered in the literature. Our main findings are summarized below.

\begin{itemize}
	\item In Sec.~\ref{sec:wakes} we detail our theoretical model of HDM wakes, which improves on previous approaches in two main directions. First, we provide fitting formulas for the ratio of neutrino and CDM transfer functions that only depend on the neutrino free-streaming scale. These are used to build a broader effective model of HDM wakes where the present-day abundance and free-streaming scales of the HDM species are uncoupled and treated as free parameters (while in the neutrino case they are fixed by the value of the neutrino mass). Second, our framework is the first to fully self-consistently model the effects of HDM wakes in the large-scale structure in terms of the bispectrum in Eq.~(\ref{eq:bispectrum}), while working for all observation scales and for arbitrary values of the effective free-streaming length (under the assumption of the squeezed limit).
	\item In Sec.~\ref{sec:detection} we discuss detection prospects, assuming cross-correlation measurements of galaxy positions and galaxy lensing. We begin in Sec.~\ref{sec:3d} with the optimistic scenario of 3D maps of the total matter field, in which case HDM wakes are observable with a billion galaxies at an effective redshift $z_{\textrm{eff}}=1$, provided that $k_{\textrm{fs},0} \gtrsim 0.03\textrm{Mpc}^{-1}$ as shown in Fig.~\ref{fig:3d_scale} (for $f_{\textrm{h}}=0.01$, in general $\textrm{SNR} \propto f_{\textrm{h}}$ as long as $f_{\textrm{h}} \ll 1$, and $\textrm{SNR} \propto \sqrt{\bar{n}_{\textrm{g}}}$ the root of the total galaxy number density). We then move on, in Sec.~\ref{sec:2d}, to the more realistic scenario of a 2D map of the total matter field, inspired by galaxy lensing, while also accounting for the noise due to the intrinsic ellipticity dispersion of weak lensing sources. We find significant penalties, in terms of signal-to-noise ratio loss, such that HDM wakes are unlikely to be ever observed in this set-up (see Tab.~\ref{table:stn_tab}). We finish in Sec.~\ref{sec:discussion} with a discussion of our results, including potential alternative ideas to detect Warm Dark Matter (WDM) with astrophysical observations. An interesting possibility is the search for WDM wakes on small-scale substructures, leveraging on the observation that the signal-to-noise increases with the effective free-streaming scale $k_{\textrm{fs}}$, as shown in Fig.~\ref{fig:3d_scale}, and that the signal-to-noise peaks at $k_{\textrm{fs}}$, as shown in Fig.~\ref{fig:scale_dependence}.
\end{itemize}

The search for HDM wakes opens up a new avenue to probe neutrino masses, and hot subcomponents of the dark matter more broadly, with cosmological observations. This is especially relevant in light of new data releases and theoretical developments, which highlight the fact that a cosmological measurement of neutrino masses (or of HDM properties more broadly) via standard techniques is very sensitive to background cosmological parameters, and hence requires precise control over the background expansion history. The exploration of additional signatures that can not be mimicked by changes to the background expansion, such as HDM wakes, is then very well motivated and has the potential to reveal new physics beyond the standard model when combined with traditional observables.   

\acknowledgments
We thank the members of the Dark Universe Science Center at the University of Washington for many helpful discussions over the past few years. CBSN thanks Matt McQuinn, John Franklin Crenshaw, Delon Shen, Matthew Johnson and Emanuele Castorina for helpful discussions. This work is supported by the Department of Energy grant DE-SC0011637, and the Dr. Ann Nelson Endowed Professorship of Physics. Research at Perimeter Institute is supported in part by the Government of Canada through the Department of Innovation, Science and Economic Development Canada and by the Province of Ontario through the Ministry of Economic Development, Job Creation and Trade. All the numerical calculations and plots in this paper were made with Mathematica \cite{Mathematica}. This version of the paper corrects an error in the numerical computation of the HDM--CDM cross power spectrum present in previous versions, which affected the signal-to-noise forecasts; we thank John Houghteling for bringing it to our attention. The authors used Claude Sonnet 5 (Anthropic) to assist in correcting this error.

\appendix 
 
\section{The short-scale limit of the neutrino dipole}
\label{sec:app1}

We show that Eq.~(\ref{eq:short_scale_dipole}) follows from Eqs.~(\ref{eq:cross_ps_monopole}) and (\ref{eq:cross_ps_dipole}) in the limit of $k \gg k_{\textrm{fs}}$. The physical reason why the equations simplify in this limit is that the collisionless dynamics on short scales approach a static limit, which is only sensitive to a snapshot of the underlying gravitational potential at the final time. This is manifested in the math as follows. Due to the appearance of the spherical Bessel function $j_{0}(xy)$ in the definition of $\Delta(x)$ in Eq.~(\ref{eq:aux_function}), this function approaches unity at $x \lesssim 1$, and it decays to zero as $x \gg 1$. Since $x=k(T_{\nu,0}/m_{\nu})(\eta-\eta')$, a large wavenumber $k$ needs to be compensated by a small $\eta-\eta'$ such that $x \lesssim 1$, and hence $\eta' \to \eta$ in the limit $k \to \infty$. From this we learn that Eq.~(\ref{eq:cross_ps_monopole}) simplifies to
\begin{equation}
\label{eq:cross_ps_monopole_highk}
	 P_{\textrm{ch}}(\eta,k) \underset{k\gg k_{\textrm{fs}}}{=} \frac{3}{2} \Omega_{\textrm{m}} H_{0}^{2} \ a(\eta) P_{\textrm{cc}}(\eta,k) \int_{0}^{\eta} d\eta' (\eta-\eta') \Delta\left(k\frac{T_{\nu,0}}{m_{\nu}}(\eta-\eta')\right) \,.
\end{equation}
 It then follows from Eq.~(\ref{eq:aux_function}), after integrating over superconformal time 
 \begin{equation}
 \label{eq:aux_calc_1}
 \begin{split}
 	\int_{0}^{\eta} d\eta' (\eta-\eta') \Delta\left(k\frac{T_{\nu,0}}{m_{\nu}}(\eta-\eta')\right) & = \frac{m_{\nu}^{2}}{k^{2}T_{\nu,0}^{2}} \int_{0}^{\infty} \frac{dy}{3\zeta(3)} f_{\textrm{FD}}(y) \left[1-\cos\left(k\frac{T_{\nu,0}}{m_{\nu}}\eta y \right)\right] \\ & \underset{k\gg k_{\textrm{fs}}}{=} \left(\frac{3}{2} \Omega_{\textrm{m}} H_{0}^{2}\right)^{-1} \frac{1}{a} \left(\frac{k_{\textrm{fs}}}{k}\right)^{2} \,,
 \end{split}
 \end{equation} 
where we used Eq.~(\ref{eq:free_streaming}). The substitution of Eq.~(\ref{eq:aux_calc_1}) into Eq.~(\ref{eq:cross_ps_monopole_highk}) yields
\begin{equation}
\label{eq:highk_nu}
	 P_{\textrm{ch}}(k) \underset{k\gg k_{\textrm{fs}}}{=} \left(\frac{k_{\textrm{fs}}}{k}\right)^{2} P_{\textrm{cc}}(k) \,,
\end{equation}
reproducing a well-known result in the literature \cite{Chen:2020kxi, Nascimento:2023psl}. We may now carry out a very similar exercise with the dipole in Eq.~(\ref{eq:cross_ps_dipole}). First notice that, from Eq.~(\ref{eq:delta_displacement})
\begin{equation}
\label{eq:rel_displacement}
	\Delta \vec{\psi}_{\textrm{rel}}(\eta,\eta') = \int_{\eta'}^{\eta} d\eta'' a(\eta'') \vec{v}_{\textrm{rel}}(\eta'') \approx a(\eta) \vec{v}_{\textrm{rel}}(\eta) (\eta-\eta') \,,
\end{equation}
in the limit of $\eta' \to \eta$. It then follows for the short-scale limit of Eq.~(\ref{eq:cross_ps_dipole})
\begin{equation}
\label{eq:cross_ps_dipole_short}
\small
	\hat{P}^{\textrm{dip}}_{\textrm{ch}}(\eta,\vec{k}) \underset{k\gg k_{\textrm{fs}}}{=} - \frac{3}{2} \Omega_{\textrm{m}} H_{0}^{2} \ a^{2}(\eta) P_{\textrm{cc}}(\eta,k) \ i\vec{k} \cdot \vec{v}_{\textrm{rel}}(\eta) \int_{0}^{\eta} d\eta' (\eta-\eta')^{2} \Delta\left(k\frac{T_{\nu,0}}{m_{\nu}}(\eta-\eta')\right)  \,.
\end{equation}
 We once again use Eq.~(\ref{eq:aux_function}) and integrate over superconformal time to obtain
 \begin{equation}
 	\label{eq:aux_calc_2}
 	\begin{split}
 		\int_{0}^{\eta} d\eta' (\eta-\eta')^{2} \Delta\left(k\frac{T_{\nu,0}}{m_{\nu}}(\eta-\eta')\right) & = \frac{m_{\nu}^{3}}{k^{3}T_{\nu,0}^{3}} \int_{0}^{\infty} \frac{dy}{3\zeta(3)} f_{\textrm{FD}}(y) \left[\frac{\sin\left(k\frac{T_{\nu,0}}{m_{\nu}} \eta y \right)}{y}-k\frac{T_{\nu,0}}{m_{\nu}} \eta \cos\left(k\frac{T_{\nu,0}}{m_{\nu}}\eta y \right)\right] \\ & \underset{k\gg k_{\textrm{fs}}}{=} \left(\frac{3}{2} \Omega_{\textrm{m}} H_{0}^{2}\right)^{-1} \frac{\gamma}{k\sigma_{\nu}a^2} \left(\frac{k_{\textrm{fs}}}{k}\right)^{2} \,, 
 \end{split}
 \end{equation} 
where we used Eqs.~(\ref{eq:free_streaming}) and (\ref{eq:nu_velocity}). The substitution of Eq.~(\ref{eq:aux_calc_2}) into Eq.~(\ref{eq:cross_ps_dipole_short}) yields
\begin{equation}
\label{eq:highk_nu_dipole}
	\hat{P}_{\textrm{ch}}^{\textrm{dip}}(k) \underset{k\gg k_{\textrm{fs}}}{=} -\gamma  \frac{i\vec{k}}{k} \cdot \frac{\vec{v}_{\textrm{rel}}}{\sigma_{\nu}} \left(\frac{k_{\textrm{fs}}}{k}\right)^{2} P_{\textrm{cc}}(k) \,.
\end{equation}
We can finally combine Eqs.~(\ref{eq:highk_nu}) and (\ref{eq:highk_nu_dipole}) to arrive at
\begin{equation}
	\label{eq:short_scale_dipole_app}
	\hat{P}^{\textrm{dip}}_{\textrm{ch}}(\vec{k}) \underset{k\gg k_{\textrm{fs}}}{=} -\gamma \frac{i\vec{k}}{k} \cdot \frac{\vec{v}_{\textrm{rel}}}{\sigma_{\nu}} P_{\textrm{ch}}(k) \,.
\end{equation}

\section{Quadratic estimators and signal-to-noise}
\label{sec:app2} 

Our goal is to derive the unbiased optimal quadratic estimators for relative velocity reconstructions. Starting with the 3D set-up, the generic form of the estimator is introduced in Eq.~(\ref{eq:3d_estimator}), which we repeat here for the reader's convenience 
\begin{equation}
\label{eq:3d_estimator_app}
	\hat{\phi}_{\textrm{rel}}(\vec{K})  = N(K) \int \frac{d^3\vec{k}}{(2\pi)^{3}} W_{A}(\vec{k},\vec{k'}) \delta_{\textrm{m}}(\vec{k})\delta_{\textrm{g}}(\vec{k}') \,.
\end{equation}
We will first impose the requirement of an unbiased estimator, i.e. $\langle \hat{\phi}_{\textrm{rel}}(\vec{K}) \rangle = \phi_{\textrm{rel}}(\vec{K})$. From the antisymmetry of $W_{A}(\vec{k},\vec{k'})$, we need to evaluate
\begin{equation}
\label{eq:unbiased_estimator}
\begin{split}
	\frac{1}{2} & \left\langle \left[ \delta_{\textrm{m}}(\vec{k})\delta_{\textrm{g}}(\vec{k}') - \delta_{\textrm{g}}(\vec{k})\delta_{\textrm{m}}(\vec{k}') \right] \right\rangle = \frac{1}{2} f_{\textrm{h}}b \left\langle \left[ \delta_{\textrm{h}}(\vec{k})\delta_{\textrm{c}}(\vec{k}') - \delta_{\textrm{c}}(\vec{k})\delta_{\textrm{h}}(\vec{k}') \right] \right\rangle \\ & = \frac{1}{2} f_{\textrm{h}}b \frac{\gamma}{\sigma_{\nu}} \  T_{\textrm{dip}}(k) (\hat{k}\cdot \hat{K}) \phi_{\textrm{rel}}(\vec{K}) P_{\textrm{ch}}(k) \,,
\end{split}
\end{equation}
where we used Eq.~(\ref{eq:pre_squeezed_bispectrum}). It then follows that
\begin{equation}
\label{eq:normalization}
	\left[N(K)\right]^{-1} = f_{\textrm{h}}b \frac{\gamma}{\sigma_{\nu}} \int \frac{d^{3}\vec{k}}{(2\pi)^{3}} W_{A}(\vec{k},\vec{k'}) (\hat{k}\cdot \hat{K}) T_{\textrm{dip}}(k) P_{\textrm{ch}}(k) \,,
\end{equation}
fixing the normalization in terms of the filter function. We next compute the covariance associated to the estimator. We obtain from Eq.~(\ref{eq:3d_estimator_app})
\begin{equation}
\label{eq:covariance_1}
	\left\langle \hat{\phi}_{\textrm{rel}}(\vec{K}) \hat{\phi}_{\textrm{rel}}(\vec{Q}) \right\rangle = N(K)N(Q) \int \frac{d^{3}\vec{k}}{(2\pi)^{3}} \int \frac{d^{3}\vec{q}}{(2\pi)^{3}} W_{A}(\vec{k},\vec{k'}) W_{A}(\vec{q},\vec{q'}) \langle \delta_{\textrm{m}}(\vec{k})\delta_{\textrm{g}}(\vec{k'}) \delta_{\textrm{m}}(\vec{q}) \delta_{\textrm{g}}(\vec{q'})\rangle \,.
\end{equation}
Using Wick's theorem to compute the Gaussian contribution to the covariance matrix, and the antisymmetry of $W_{A}(\vec{k},\vec{k'})$, one obtains
\begin{equation}
\label{eq:covariance_2}
	\left\langle \hat{\phi}_{\textrm{rel}}(\vec{K}) \hat{\phi}_{\textrm{rel}}(\vec{Q}) \right\rangle = (2\pi)^{3}\delta^{(3)}(\vec{K}+\vec{Q}) \textrm{Cov}(K) \,,
\end{equation}
where
\begin{equation}
\label{eq:covariance_3}
	\textrm{Cov}(K) = [N(K)]^{2} \int\frac{d^{3}\vec{k}}{(2\pi)^{3}} \left[W_{A}(\vec{k},\vec{k'})\right]^{2} P_{\textrm{gg}}(k)P_{\textrm{mm}}(k)\left[1-r^{2}(k)\right] \,.
\end{equation}
We add a generic white noise to the total matter field and shot-noise to the galaxy power spectrum [see Eq.~(\ref{eq:galaxy_power}) and lines of text below]. Also, $r(k)$ stands for the galaxy-matter correlation coefficient as defined in Eq.~(\ref{eq:3d_correlation}). The next step is to impose the requirement of an optimal estimator, which involves taking a functional derivative of  Eq.~(\ref{eq:covariance_3}) with respect to the filter and setting it to zero, in order to find the filter that minimizes the covariance. Using Eq.~(\ref{eq:normalization}), this yields after a straightforward calculation 
\begin{equation}
\label{eq:optimal_filter}
	W(\vec{k},\vec{k'}) = f_{\textrm{h}}b \frac{\gamma}{\sigma_{\nu}} \frac{(\hat{k}\cdot \hat{K})T_{\textrm{dip}}(k) P_{\textrm{ch}}(k)}{P_{\textrm{gg}}(k)P_{\textrm{mm}}(k)\left[1-r^{2}(k)\right]} \,.
\end{equation}
We may now substitute Eq.~(\ref{eq:optimal_filter}) into Eqs.~(\ref{eq:normalization}) and (\ref{eq:covariance_3}) to find
\begin{equation}
\label{eq:covariance_4}
	\textrm{Cov}(K)^{-1} = N(K)^{-1} = \frac{1}{3} f_{\textrm{h}}^2 b^{2} \frac{\gamma^{2}}{\sigma_{\nu}^2} \int \frac{k^2 dk}{2\pi^{2}} \frac{[T_{\textrm{dip}}(k)P_{\textrm{ch}}(k)]^2}{P_{\textrm{gg}}(k)P_{\textrm{mm}}(k)\left[1-r^{2}(k)\right]} \,,
\end{equation}
corresponding exactly to Eq.~(\ref{eq:3d_covariance}) in the main text.

The derivation of the optimal unbiased estimator in the two-dimensional case, and the associated cumulative signal-to-noise, is entirely analogous to the calculations done here within the three-dimensional set-up so we will omit most of the details. The 2D analog of Eq.~(\ref{eq:covariance_4}) reads
\begin{equation}
\label{eq:2d_covariance_app}
	N(L)^{-1} = \frac{1}{2} f_{\textrm{h}}^2 \int \frac{\ell d\ell}{2\pi} \alpha_{l,L}^{2} \frac{\left[P_{\textrm{ch}}(\ell)\right]^2}{P_{\textrm{gg}}(\ell)P_{\textrm{mm}}(\ell)\left[1-r^{2}(\ell)\right]} \,,
\end{equation}
where the various ingredients entering Eq.~(\ref{eq:2d_covariance_app}) are defined in Sec.~\ref{sec:2d}. We are now ready to write down the formula for the 2D cumulative signal-to-noise ratio
\begin{equation}
\begin{split}
	\label{eq:snr_2d_app}
	\textrm{SNR}^{2}_{\textrm{2d}} & = \Omega \int \frac{d^2\vec{L}}{(2\pi)^{2}} \frac{P_{\phi_{\textrm{rel}}\phi_{\textrm{rel}}}(L)}{N(L)} \\ & = \frac{1}{2\pi} f_{\textrm{sky}} f_{\textrm{h}}^{2} \int LdL \int \ell d\ell \ \alpha_{\ell,L}^{2} \frac{ \left[P_{\textrm{ch}}(\ell)\right]^{2} P_{\phi_{\textrm{rel}}\phi_{\textrm{rel}}}(L) }{P_{\kappa\kappa}(\ell)P_{\textrm{g} \textrm{g}}(\ell)[1-r^{2}(\ell)]} \,,
\end{split}
\end{equation}
where $\Omega = 4\pi f_{\textrm{sky}}$ is the total solid angle covered by the survey, matching Eq.~(\ref{eq:2d_stn}) in the main text. 

\section{Optimal combination of redshift bins}
\label{sec:app3} 

The tomographic approach we take is to optimally combine the $N_{\textrm{z-bins}}$ redshift bins into a single galaxy and relative velocity potential two-dimensional fields, inspired by techniques introduced in \cite{Sherwin:2015baa, Schmittfull:2017ffw}. We now set-up this optimization problem and present its solution in terms of the coefficients $c^{\ell}_{i}$ and $d^{\ell,L}_{i}$, defined as
\begin{equation}
	\label{eq:optimal_1_app}
	\delta_{\textrm{g}} = \sum_{i=1}^{N_{\textrm{z-bins}}} c^{\ell}_{i} 	\delta_{\textrm{g}_{i}} \,,
\end{equation}
and
\begin{equation}
	\label{eq:optimal_2_app}
	\phi_{\textrm{rel}} = \sum_{i=1}^{N_{\textrm{z-bins}}} d^{\ell,L}_{i} 	\delta_{\textrm{g}_{i}} \,.
\end{equation}
These correspond to Eqs.~(\ref{eq:optimal_1}) and (\ref{eq:optimal_2}) in the main text, which we repeat here for the readers convenience. Moving forward we will drop the multipole dependence of the coefficients (which are redshift independent by construction) for simplicity of notation. We first maximize the correlation coefficient squared
\begin{equation}
	\label{eq:2d_corr_coefficient_app}
	r^{2}(l) = \frac{	P_{\textrm{g}\kappa}^{2}(\ell)}{P_{\textrm{gg}}(\ell) P_{\kappa\kappa}(\ell)} \,,
\end{equation}
to obtain the coefficients $c_{i}$. The details of this procedure can be found in Appendix A of \cite{Sherwin:2015baa}, where the following formula is derived (for non-overlapping redshift bins as in our present case)
\begin{equation}
	\label{eq:sol_coef_1}
	c_{\alpha} = \frac{P_{\kappa \textrm{g}_{\alpha}}}{P_{\textrm{g}_{\alpha}\textrm{g}_{\alpha}}} \,.
\end{equation}
From this it follows that
\begin{equation}
	\label{eq:optimal_cross}
	P_{\textrm{g}\kappa} = \sum_{i} c_{i} P_{\kappa \textrm{g}_{i}} = \sum_{i} \frac{P_{\kappa \textrm{g}_{i}}^{2}}{P_{\textrm{g}_{i}\textrm{g}_{i}}} \,,
\end{equation}
and
\begin{equation}
	\label{eq:optimal_auto}
	P_{\textrm{g} \textrm{g}} = \sum_{i} c_{i}^{2} P_{\textrm{g}_{i} \textrm{g}_{i}} = \sum_{i} \frac{P_{\kappa \textrm{g}_{i}}^{2}}{P_{\textrm{g}_{i}\textrm{g}_{i}}} \equiv P_{\textrm{g}\kappa} \,.
\end{equation}
The substitution of Eqs.~(\ref{eq:optimal_cross}) and (\ref{eq:optimal_auto}) into Eq.~(\ref{eq:2d_corr_coefficient_app}) yields
\begin{equation}
	\label{eq:optimal_correlation}
	r^{2} = \sum_{i} r_{i}^{2} = \sum_{i} \frac{P_{\kappa \textrm{g}_{i}}^{2}}{P_{\kappa \kappa}P_{\textrm{g}_{i}\textrm{g}_{i}}} \,,
\end{equation}
such that the optimal correlation coefficient is the sum in quadrature of the correlation coefficients from different redshift bins. This is plotted in Fig.~\ref{fig:optimal_correlation} for $N_{\textrm{z-bins}}=6$. Finally, combining Eqs.~(\ref{eq:optimal_cross}), (\ref{eq:optimal_auto}) and (\ref{eq:optimal_correlation}) gives
\begin{equation}
	\label{eq:optimal_ps}
	P_{\textrm{g} \textrm{g}} =  P_{\textrm{g}\kappa} = r^{2} P_{\kappa\kappa} \,.
\end{equation}

Let us now move on to obtaining the $d_{i}$ coefficients in Eq.~(\ref{eq:optimal_2_app}). We will do so by taking the derivatives of the signal-to-noise with respect to these coefficients, and setting them to zero. From Eq.~(\ref{eq:2d_stn})
\begin{equation}
\label{eq:str_scaling}
	\textrm{SNR}^{2} \propto \alpha_{\ell,L}^{2} P_{\phi_{\textrm{rel}}\phi_{\textrm{rel}}} \propto \frac{\beta_{\ell,L}^2}{P_{\phi_{\textrm{rel}}\phi_{\textrm{rel}}}} \,,
\end{equation}
where based on Eq.~(\ref{eq:2d_coefficient}), we write
\begin{equation}
\label{eq:beta_coefficient}
	\alpha_{\ell,L} = \frac{\beta_{\ell,L}}{P_{\textrm{ch}}(\ell) P_{\phi_{\textrm{rel}}\phi_{\textrm{rel}}}(L)} \,,
\end{equation}
with
\begin{equation}
	\label{eq:beta_coefficient_2}
	\small
	\beta_{\ell, L} =  \int dz \frac{H^{2}(z)}{x_{\parallel}^{4}(z)} W_{\textrm{g}}(z) W_{\kappa}(z) W_{\phi_{\textrm{rel}}}(z) \frac{\gamma}{\sigma_{\nu}(z)} T_{\textrm{dip}}\left(z, \frac{\ell}{x_{\parallel}(z)}\right) P_{\textrm{ch}}\left(z, \frac{\ell}{x_{\parallel}(z)}\right) P_{\phi_{\textrm{rel}}\phi_{\textrm{rel}}}\left(z, \frac{L}{x_{\parallel}(z)}\right) \,.
\end{equation}
The linear combinations in Eqs.~(\ref{eq:optimal_1_app}) and (\ref{eq:optimal_2_app}) can be implemented by letting $W_{\textrm{g}} \to \sum_{i} c_{i}W_{\textrm{g},i}$ and $W_{\phi_{\textrm{rel}}} \to \sum_{i} d_{i}W_{\textrm{g},i}$. It then follows from the second line of Eq.~(\ref{eq:2d_ps}) and Eq.~(\ref{eq:beta_coefficient_2}) that
\begin{equation}
	\label{eq:str_scaling_2}
	\textrm{SNR}_{\textrm{2d}}^{2} \propto \frac{\beta_{\ell,L}^2}{P_{\phi_{\textrm{rel}}\phi_{\textrm{rel}}}} = \frac{\left( \sum_{i} c_{i} d_{i} \beta_{\ell, L}^{i} \right)^2}{\sum_{i} d_{i}^{2} P_{\phi_{\textrm{rel}}\phi_{\textrm{rel}}}^{i}} \,,
\end{equation}
where $\beta_{\ell, L}^{i}$ and $P_{\phi_{\textrm{rel}}\phi_{\textrm{rel}}}^{i}$ can be computed following the replacements $W_{\textrm{g}} \to W_{\textrm{g},i}$ and $W_{\phi_{\textrm{rel}}} \to W_{\textrm{g},i}$. We may now take the derivative of the logarithm of the signal-to-noise with respect to $d_{\alpha}$ and set it to zero as follows
\begin{equation}
\label{eq:stn_partial}
	\frac{\partial}{\partial d_{\alpha}} \log \textrm{SNR}_{\textrm{2d}}^{2} = \frac{2c_{\alpha} \beta_{\ell,L}^{\alpha}}{\sum_{i} c_{i} d_{i} \beta_{\ell,L}^{i}} - \frac{2d_{\alpha} P_{\phi_{\textrm{rel}}\phi_{\textrm{rel}}}^{\alpha}}{\sum_{i}d_{i}^{2} P_{\phi_{\textrm{rel}}\phi_{\textrm{rel}}}^{i}} = 0 \implies d_{\alpha} \propto \frac{\beta_{\ell,L}^{\alpha}}{P_{\phi_{\textrm{rel}}\phi_{\textrm{rel}}}^{\alpha}} c_{\alpha} \,.
\end{equation}
An arbitrary rescaling of the coefficients $d_{i}$ by a fixed normalization factor does not change the signal-to-noise, as is clear from Eq.~(\ref{eq:str_scaling_2}). Hence, the proportionality factor in Eq.~(\ref{eq:stn_partial}) is arbitrary. We can then write, combining Eqs.~(\ref{eq:beta_coefficient}) and (\ref{eq:stn_partial})
\begin{equation}
\label{eq:sol_coef_2}
	d_{\alpha} \propto \alpha_{\ell,L}^{\alpha} P_{\textrm{ch}}^{\alpha} c_{\alpha} \equiv \alpha_{\ell,L}^{\alpha} \frac{P_{\textrm{ch}}^{\alpha}}{P_{\textrm{ch}}}  c_{\alpha}  \,,
\end{equation}
where we inserted a factor of
\begin{equation}
\label{eq:optimal_cross_2}
	P_{\textrm{ch}} = \sum_{i} c_{i} P_{\textrm{ch}}^{i} \,,
\end{equation}
in the denominator to fix the normalization. The motivation for this particular choice is the following. Combining Eqs.~(\ref{eq:beta_coefficient}) and (\ref{eq:sol_coef_2}) yields
\begin{equation}
\label{eq:alpha_optimal}
	\alpha_{\ell,L} = \frac{\sum_{i} c_{i}d_{i} \beta_{\ell,L}^{i}}{P_{\textrm{ch}} P_{\phi_{\textrm{rel}}\phi_{\textrm{rel}}}} = \frac{1}{P_{\phi_{\textrm{rel}}\phi_{\textrm{rel}}}} \sum_{i} \left(\alpha_{\ell,L}^{\alpha} \frac{P_{\textrm{ch}}^{\alpha}}{P_{\textrm{ch}}}  c_{\alpha}\right)^{2} P_{\phi_{\textrm{rel}}\phi_{\textrm{rel}}}^{i} \equiv \frac{1}{P_{\phi_{\textrm{rel}}\phi_{\textrm{rel}}}} \sum_{i} d_{i}^{2} P_{\phi_{\textrm{rel}}\phi_{\textrm{rel}}}^{i} = 1 \,.
\end{equation}
\bibliography{measuring_wakes.bib}

\begin{thebibliography}{100}

\bibitem{SimonsObservatory:2025wwn}
M.~Abitbol et~al.
\newblock {The Simons Observatory: Science Goals and Forecasts for the Enhanced Large Aperture Telescope}.
\newblock 3 2025.

\bibitem{CMB-S4:2022ght}
Kevork Abazajian et~al.
\newblock {Snowmass 2021 CMB-S4 White Paper}.
\newblock 3 2022.

\bibitem{DESI:2025fxa}
M.~Abdul~Karim et~al.
\newblock {Data Release 1 of the Dark Energy Spectroscopic Instrument}.
\newblock 3 2025.

\bibitem{Ferraro:2022cmj}
Simone Ferraro, Noah Sailer, Anze Slosar, and Martin White.
\newblock {Snowmass2021 Cosmic Frontier White Paper: Cosmology and Fundamental Physics from the three-dimensional Large Scale Structure}.
\newblock 3 2022.

\bibitem{Euclid:2024yrr}
Y.~Mellier et~al.
\newblock {Euclid. I. Overview of the Euclid mission}.
\newblock {\em Astron. Astrophys.}, 697:A1, 2025.

\bibitem{crill2020spherex}
Brendan~P Crill, Michael Werner, Rachel Akeson, Matthew Ashby, Lindsey Bleem, James~J Bock, Sean Bryan, Jill Burnham, Joyce Byunh, Tzu-Ching Chang, et~al.
\newblock Spherex: Nasa's near-infrared spectrophotometric all-sky survey.
\newblock In {\em Space Telescopes and Instrumentation 2020: Optical, Infrared, and Millimeter Wave}, volume 11443, pages 61--77. SPIE, 2020.

\bibitem{Mao:2022fyx}
Yao-Yuan Mao et~al.
\newblock {Snowmass2021: Vera C. Rubin Observatory as a Flagship Dark Matter Experiment}.
\newblock 3 2022.

\bibitem{Dore:2019pld}
Olivier Dor\'e et~al.
\newblock {WFIRST: The Essential Cosmology Space Observatory for the Coming Decade}.
\newblock 4 2019.

\bibitem{Gong:2025ecr}
Yan Gong et~al.
\newblock {Future Cosmology: New Physics and Opportunity from the China Space Station Telescope (CSST)}.
\newblock {\em Sci. China Phys. Mech. Astron.}, 68:280402, 2025.

\bibitem{Yu:2025rez}
Tien-Tien Yu.
\newblock {2024 TASI Lectures: A Dark Matter Primer}.
\newblock 6 2025.

\bibitem{Cirelli:2024ssz}
Marco Cirelli, Alessandro Strumia, and Jure Zupan.
\newblock {Dark Matter}.
\newblock 6 2024.

\bibitem{Peters:2023asu}
Fabian~Hervas Peters, Aurel Schneider, Jozef Bucko, Sambit~K. Giri, and Gabriele Parimbelli.
\newblock {Constraining hot dark matter sub-species with weak lensing and the cosmic microwave background radiation}.
\newblock {\em Astron. Astrophys.}, 687:A161, 2024.

\bibitem{Garcia-Gallego:2025kiw}
Olga Garcia-Gallego, Vid Ir\v{s}i\v{c}, Martin~G. Haehnelt, Matteo Viel, and James~S. Bolton.
\newblock {Constraining Mixed Dark Matter models with high redshift Lyman-alpha forest data}.
\newblock 4 2025.

\bibitem{Euclid:2024pwi}
J.~Lesgourgues et~al.
\newblock {Euclid preparation - LVI. Sensitivity to non-standard particle dark matter models}.
\newblock {\em Astron. Astrophys.}, 693:A249, 2025.

\bibitem{Dror:2024ibf}
Jeff~A. Dror, Pearl Sandick, Barmak Shams Es~Haghi, and Fengwei Yang.
\newblock {Indirect detection of hot dark matter}.
\newblock {\em JHEP}, 05:127, 2025.

\bibitem{Xu:2021rwg}
Weishuang~Linda Xu, Julian~B. Mu\~noz, and Cora Dvorkin.
\newblock {Cosmological constraints on light but massive relics}.
\newblock {\em Phys. Rev. D}, 105(9):095029, 2022.

\bibitem{Dayal:2023nwi}
Pratika Dayal and Sambit~K. Giri.
\newblock {Warm dark matter constraints from the JWST}.
\newblock {\em Mon. Not. Roy. Astron. Soc.}, 528(2):2784--2789, 2024.

\bibitem{Verdiani:2025jcf}
Francesco Verdiani, Emanuele Castorina, Ennio Salvioni, and Emiliano Sefusatti.
\newblock {The Effective Field Theory of Large Scale Structure for Mixed Dark Matter Scenarios}.
\newblock 7 2025.

\bibitem{Boyarsky:2008xj}
Alexey Boyarsky, Julien Lesgourgues, Oleg Ruchayskiy, and Matteo Viel.
\newblock {Lyman-alpha constraints on warm and on warm-plus-cold dark matter models}.
\newblock {\em JCAP}, 05:012, 2009.

\bibitem{deSalas:2020pgw}
P.~F. de~Salas, D.~V. Forero, S.~Gariazzo, P.~Mart\'\i{}nez-Mirav\'e, O.~Mena, C.~A. Ternes, M.~T\'ortola, and J.~W.~F. Valle.
\newblock {2020 global reassessment of the neutrino oscillation picture}.
\newblock {\em JHEP}, 02:071, 2021.

\bibitem{Esteban:2020cvm}
Ivan Esteban, M.~C. Gonzalez-Garcia, Michele Maltoni, Thomas Schwetz, and Albert Zhou.
\newblock {The fate of hints: updated global analysis of three-flavor neutrino oscillations}.
\newblock {\em JHEP}, 09:178, 2020.

\bibitem{Gonzalez-Garcia:2021dve}
Maria~Concepcion Gonzalez-Garcia, Michele Maltoni, and Thomas Schwetz.
\newblock {NuFIT: Three-Flavour Global Analyses of Neutrino Oscillation Experiments}.
\newblock {\em Universe}, 7(12):459, 2021.

\bibitem{KATRIN:2024cdt}
Max Aker et~al.
\newblock {Direct neutrino-mass measurement based on 259 days of KATRIN data}.
\newblock {\em Science}, 388(6743):adq9592, 2025.

\bibitem{Planck:2018vyg}
N.~Aghanim et~al.
\newblock {Planck 2018 results. VI. Cosmological parameters}.
\newblock {\em Astron. Astrophys.}, 641:A6, 2020.
\newblock [Erratum: Astron.Astrophys. 652, C4 (2021)].

\bibitem{ACT:2025tim}
Erminia Calabrese et~al.
\newblock {The Atacama Cosmology Telescope: DR6 Constraints on Extended Cosmological Models}.
\newblock 3 2025.

\bibitem{DESI:2024hhd}
A.~G. Adame et~al.
\newblock {DESI 2024 VII: Cosmological Constraints from the Full-Shape Modeling of Clustering Measurements}.
\newblock 11 2024.

\bibitem{Ivanov:2024jtl}
Mikhail~M. Ivanov, Michael~W. Toomey, and Naim~G\"oksel Kara\c{c}ayl\i{}.
\newblock {Fundamental Physics with the Lyman-Alpha Forest: Constraints on the Growth of Structure and Neutrino Masses from SDSS with Effective Field Theory}.
\newblock {\em Phys. Rev. Lett.}, 134(9):091001, 2025.

\bibitem{DESI:2025ejh}
W.~Elbers et~al.
\newblock {Constraints on Neutrino Physics from DESI DR2 BAO and DR1 Full Shape}.
\newblock 3 2025.

\bibitem{Green:2021xzn}
Daniel Green and Joel Meyers.
\newblock {Cosmological Implications of a Neutrino Mass Detection}.
\newblock 11 2021.

\bibitem{Escudero:2020ped}
Miguel Escudero, Jacobo Lopez-Pavon, Nuria Rius, and Stefan Sandner.
\newblock {Relaxing Cosmological Neutrino Mass Bounds with Unstable Neutrinos}.
\newblock {\em JHEP}, 12:119, 2020.

\bibitem{daFonseca:2023ury}
Vitor da~Fonseca, Tiago Barreiro, and Nelson~J. Nunes.
\newblock {Relaxing cosmological constraints on current neutrino masses}.
\newblock {\em Phys. Rev. D}, 109(6):063517, 2024.

\bibitem{FrancoAbellan:2021hdb}
Guillermo Franco~Abell\'an, Zackaria Chacko, Abhish Dev, Peizhi Du, Vivian Poulin, and Yuhsin Tsai.
\newblock {Improved cosmological constraints on the neutrino mass and lifetime}.
\newblock {\em JHEP}, 08:076, 2022.

\bibitem{Hannestad:2005gj}
Steen Hannestad.
\newblock {Neutrino masses and the dark energy equation of state - Relaxing the cosmological neutrino mass bound}.
\newblock {\em Phys. Rev. Lett.}, 95:221301, 2005.

\bibitem{DESI:2025fii}
K.~Lodha et~al.
\newblock {Extended Dark Energy analysis using DESI DR2 BAO measurements}.
\newblock 3 2025.

\bibitem{Reboucas:2024smm}
Jo\~ao Rebou\c{c}as, Diogo H.~F. de~Souza, Kunhao Zhong, Vivian Miranda, and Rogerio Rosenfeld.
\newblock {Investigating late-time dark energy and massive neutrinos in light of DESI Y1 BAO}.
\newblock {\em JCAP}, 02:024, 2025.

\bibitem{Racco:2024lbu}
Davide Racco, Pierre Zhang, and Henry Zheng.
\newblock {Neutrino masses from large-scale structures: Future sensitivity and theory dependence}.
\newblock {\em Phys. Dark Univ.}, 47:101803, 2025.

\bibitem{Green:2024xbb}
Daniel Green and Joel Meyers.
\newblock {Cosmological preference for a negative neutrino mass}.
\newblock {\em Phys. Rev. D}, 111(8):083507, 2025.

\bibitem{Craig:2024tky}
Nathaniel Craig, Daniel Green, Joel Meyers, and Surjeet Rajendran.
\newblock {No \ensuremath{\nu}s is Good News}.
\newblock {\em JHEP}, 09:097, 2024.

\bibitem{Loverde:2024nfi}
Marilena Loverde and Zachary~J. Weiner.
\newblock {Massive neutrinos and cosmic composition}.
\newblock {\em JCAP}, 12:048, 2024.

\bibitem{Lynch:2025ine}
Gabriel~P. Lynch and Lloyd Knox.
\newblock {What's the matter with $\Sigma m_{\nu}$?}
\newblock 3 2025.

\bibitem{Graham:2025fdt}
Peter~W. Graham, Daniel Green, and Joel Meyers.
\newblock {Dark Forces Gathering}.
\newblock 8 2025.

\bibitem{Tishue:2025zdw}
Avery~J. Tishue, Selim~C. Hotinli, Peter Adshead, Ely~D. Kovetz, and Mathew~S. Madhavacheril.
\newblock {Neutrino Mass Constraints from kSZ Tomography}.
\newblock 2 2025.

\bibitem{LoVerde:2016ahu}
Marilena LoVerde.
\newblock {Neutrino mass without cosmic variance}.
\newblock {\em Phys. Rev. D}, 93(10):103526, 2016.

\bibitem{Rogozenski:2023tse}
P.~Rogozenski, E.~Krause, and V.~Miranda.
\newblock {Modeling neutrino-induced scale-dependent galaxy clustering for photometric galaxy surveys}.
\newblock {\em JCAP}, 04:076, 2024.

\bibitem{Marques:2018ctl}
Gabriela~A. Marques, Jia Liu, Jos\'e Manuel~Zorrilla Matilla, Zolt\'an Haiman, Armando Bernui, and Camila~P. Novaes.
\newblock {Constraining neutrino mass with weak lensing Minkowski Functionals}.
\newblock {\em JCAP}, 06:019, 2019.

\bibitem{Kreisch:2018var}
Christina~D. Kreisch, Alice Pisani, Carmelita Carbone, Jia Liu, Adam~J. Hawken, Elena Massara, David~N. Spergel, and Benjamin~D. Wandelt.
\newblock {Massive Neutrinos Leave Fingerprints on Cosmic Voids}.
\newblock {\em Mon. Not. Roy. Astron. Soc.}, 488(3):4413--4426, 2019.

\bibitem{Shiveshwarkar:2020jxr}
Charuhas Shiveshwarkar, Drew Jamieson, and Marilena Loverde.
\newblock {Scale-dependent halo bias and the squeezed limit bispectrum in the presence of radiation}.
\newblock {\em Phys. Rev. D}, 103(10):103503, 2021.

\bibitem{Pal:2025hpl}
Sourav Pal, Debanjan Sarkar, Rickmoy Samanta, and Supratik Pal.
\newblock {Redshift-space galaxy bispectrum in presence of massive neutrinos: A multipole expansion approach for Euclid}.
\newblock {\em Mon. Not. Roy. Astron. Soc.}, 542:1, 2025.

\bibitem{tseliakhovich2010relative}
Dmitriy Tseliakhovich and Christopher Hirata.
\newblock Relative velocity of dark matter and baryonic fluids and the formation of the first structures.
\newblock {\em Physical Review D—Particles, Fields, Gravitation, and Cosmology}, 82(8):083520, 2010.

\bibitem{mcquinn2012impact}
Matthew McQuinn and Ryan~M O'Leary.
\newblock The impact of the supersonic baryon--dark matter velocity difference on the z~ 20 21 cm background.
\newblock {\em The Astrophysical Journal}, 760(1):3, 2012.

\bibitem{Zhu:2013tma}
Hong-Ming Zhu, Ue-Li Pen, Xuelei Chen, Derek Inman, and Yu~Yu.
\newblock {Measurement of Neutrino Masses from Relative Velocities}.
\newblock {\em Phys. Rev. Lett.}, 113:131301, 2014.

\bibitem{Zhu:2019kzb}
Hong-Ming Zhu and Emanuele Castorina.
\newblock {Measuring dark matter-neutrino relative velocity on cosmological scales}.
\newblock {\em Phys. Rev. D}, 101(2):023525, 2020.

\bibitem{Zhu:2014qma}
Hong-Ming Zhu, Ue-Li Pen, Xuelei Chen, and Derek Inman.
\newblock {Probing Neutrino Hierarchy and Chirality via Wakes}.
\newblock {\em Phys. Rev. Lett.}, 116(14):141301, 2016.

\bibitem{Okoli:2016vmd}
Chiamaka Okoli, Morag~I. Scrimgeour, Niayesh Afshordi, and Michael~J. Hudson.
\newblock {Dynamical friction in the primordial neutrino sea}.
\newblock {\em Mon. Not. Roy. Astron. Soc.}, 468(2):2164--2175, 2017.

\bibitem{Ge:2023nnh}
Shao-Feng Ge, Pedro Pasquini, and Liang Tan.
\newblock {Neutrino mass measurement with cosmic gravitational focusing}.
\newblock {\em JCAP}, 05:108, 2024.

\bibitem{Ge:2024kac}
Shao-Feng Ge and Liang Tan.
\newblock {Identifying neutrino mass ordering with cosmic gravitational focusing}.
\newblock {\em Phys. Rev. D}, 111(8):083539, 2025.

\bibitem{Ge:2025ctw}
Shao-Feng Ge and Liang Tan.
\newblock {Probing Light Dark Matter with Cosmic Gravitational Focusing}.
\newblock 9 2025.

\bibitem{Inman:2016prk}
Derek Inman, Hao-Ran Yu, Hong-Ming Zhu, J.~D. Emberson, Ue-Li Pen, Tong-Jie Zhang, Shuo Yuan, Xuelei Chen, and Zhi-Zhong Xing.
\newblock {Simulating the cold dark matter-neutrino dipole with TianNu}.
\newblock {\em Phys. Rev. D}, 95(8):083518, 2017.

\bibitem{Hernandez-Molinero:2024sds}
Beatriz Hern\'andez-Molinero, Carmelita Carbone, Raul Jimenez, and Carlos Pe\~na Garay.
\newblock {Neutrino halo profiles: HR-DEMNUni simulation analysis}.
\newblock {\em JCAP}, 09:033, 2024.

\bibitem{Nascimento:2023ezc}
Caio Nascimento and Marilena Loverde.
\newblock {Neutrino winds on the sky}.
\newblock {\em JCAP}, 11:036, 2023.

\bibitem{LoVerde:2014pxa}
Marilena LoVerde.
\newblock {Halo bias in mixed dark matter cosmologies}.
\newblock {\em Phys. Rev. D}, 90(8):083530, 2014.

\bibitem{Villaescusa-Navarro:2017mfx}
Francisco Villaescusa-Navarro, Arka Banerjee, Neal Dalal, Emanuele Castorina, Roman Scoccimarro, Raul Angulo, and David~N. Spergel.
\newblock {The imprint of neutrinos on clustering in redshift-space}.
\newblock {\em Astrophys. J.}, 861(1):53, 2018.

\bibitem{Castorina:2015bma}
Emanuele Castorina, Carmelita Carbone, Julien Bel, Emiliano Sefusatti, and Klaus Dolag.
\newblock {DEMNUni: The clustering of large-scale structures in the presence of massive neutrinos}.
\newblock {\em JCAP}, 07:043, 2015.

\bibitem{Costanzi:2013bha}
Matteo Costanzi, Francisco Villaescusa-Navarro, Matteo Viel, Jun-Qing Xia, Stefano Borgani, Emanuele Castorina, and Emiliano Sefusatti.
\newblock {Cosmology with massive neutrinos III: the halo mass function and an application to galaxy clusters}.
\newblock {\em JCAP}, 12:012, 2013.

\bibitem{Castorina:2013wga}
Emanuele Castorina, Emiliano Sefusatti, Ravi~K. Sheth, Francisco Villaescusa-Navarro, and Matteo Viel.
\newblock {Cosmology with massive neutrinos II: on the universality of the halo mass function and bias}.
\newblock {\em JCAP}, 02:049, 2014.

\bibitem{Villaescusa-Navarro:2013pva}
Francisco Villaescusa-Navarro, Federico Marulli, Matteo Viel, Enzo Branchini, Emanuele Castorina, Emiliano Sefusatti, and Shun Saito.
\newblock {Cosmology with massive neutrinos I: towards a realistic modeling of the relation between matter, haloes and galaxies}.
\newblock {\em JCAP}, 03:011, 2014.

\bibitem{Verdiani:2025znc}
Francesco Verdiani, Emilio Bellini, Chiara Moretti, Emiliano Sefusatti, Carmelita Carbone, and Matteo Viel.
\newblock {Redshift-Space Distortions in Massive Neutrinos Cosmologies}.
\newblock 3 2025.

\bibitem{marulli2011effects}
Federico Marulli, Carmelita Carbone, Matteo Viel, Lauro Moscardini, and Andrea Cimatti.
\newblock Effects of massive neutrinos on the large-scale structure of the universe.
\newblock {\em Monthly Notices of the Royal Astronomical Society}, 418(1):346--356, 2011.

\bibitem{Lesgourgues:2006nd}
Julien Lesgourgues and Sergio Pastor.
\newblock {Massive neutrinos and cosmology}.
\newblock {\em Phys. Rept.}, 429:307--379, 2006.

\bibitem{Ringwald:2004np}
Andreas Ringwald and Yvonne Y.~Y. Wong.
\newblock {Gravitational clustering of relic neutrinos and implications for their detection}.
\newblock {\em JCAP}, 12:005, 2004.

\bibitem{ali2013efficient}
Yacine Ali-Haimoud and Simeon Bird.
\newblock An efficient implementation of massive neutrinos in non-linear structure formation simulations.
\newblock {\em Monthly Notices of the Royal Astronomical Society}, 428(4):3375--3389, 2013.

\bibitem{Holm:2024zpr}
Emil~Brinch Holm, Stefan Zentarra, and Isabel~M. Oldengott.
\newblock {Local clustering of relic neutrinos: comparison of kinetic field theory and the Vlasov equation}.
\newblock {\em JCAP}, 07:050, 2024.

\bibitem{Hotinli:2023scz}
Selim~C. Hotinli, Nashwan Sabti, Jaxon North, and Marc Kamionkowski.
\newblock {Unveiling neutrino halos with CMB lensing}.
\newblock {\em Phys. Rev. D}, 108(10):103504, 2023.

\bibitem{Hu:2001kj}
Wayne Hu and Takemi Okamoto.
\newblock {Mass reconstruction with cmb polarization}.
\newblock {\em Astrophys. J.}, 574:566--574, 2002.

\bibitem{Hirata:2003ka}
Christopher~M. Hirata and Uros Seljak.
\newblock {Reconstruction of lensing from the cosmic microwave background polarization}.
\newblock {\em Phys. Rev. D}, 68:083002, 2003.

\bibitem{Maniyar:2021msb}
Abhishek~S. Maniyar, Yacine Ali-Ha\"\i{}moud, Julien Carron, Antony Lewis, and Mathew~S. Madhavacheril.
\newblock {Quadratic estimators for CMB weak lensing}.
\newblock {\em Phys. Rev. D}, 103(8):083524, 2021.

\bibitem{Smith:2018bpn}
Kendrick~M. Smith, Mathew~S. Madhavacheril, Moritz M\"unchmeyer, Simone Ferraro, Utkarsh Giri, and Matthew~C. Johnson.
\newblock {KSZ tomography and the bispectrum}.
\newblock 10 2018.

\bibitem{Deutsch:2017ybc}
Anne-Sylvie Deutsch, Emanuela Dimastrogiovanni, Matthew~C. Johnson, Moritz M\"unchmeyer, and Alexandra Terrana.
\newblock {Reconstruction of the remote dipole and quadrupole fields from the kinetic Sunyaev Zel\textquoteright{}dovich and polarized Sunyaev Zel\textquoteright{}dovich effects}.
\newblock {\em Phys. Rev. D}, 98(12):123501, 2018.

\bibitem{Hotinli:2025tul}
Selim~C. Hotinli, Kendrick~M. Smith, and Simone Ferraro.
\newblock {Velocity Reconstruction from KSZ: Measuring $f_{NL}$ with ACT and DESILS}.
\newblock 6 2025.

\bibitem{Mead:2020vgs}
Alexander Mead, Samuel Brieden, Tilman Tr\"oster, and Catherine Heymans.
\newblock {hmcode-2020: improved modelling of non-linear cosmological power spectra with baryonic feedback}.
\newblock {\em Mon. Not. Roy. Astron. Soc.}, 502(1):1401--1422, 2021.

\bibitem{blas2011cosmic}
Diego Blas, Julien Lesgourgues, and Thomas Tram.
\newblock The cosmic linear anisotropy solving system (class). part ii: Approximation schemes.
\newblock {\em Journal of Cosmology and Astroparticle Physics}, 2011(07):034, 2011.

\bibitem{Desjacques:2016bnm}
Vincent Desjacques, Donghui Jeong, and Fabian Schmidt.
\newblock {Large-Scale Galaxy Bias}.
\newblock {\em Phys. Rept.}, 733:1--193, 2018.

\bibitem{McDonald:2008sh}
Patrick McDonald and Uros Seljak.
\newblock {How to measure redshift-space distortions without sample variance}.
\newblock {\em JCAP}, 10:007, 2009.

\bibitem{Seljak:2008xr}
Uros Seljak.
\newblock {Extracting primordial non-gaussianity without cosmic variance}.
\newblock {\em Phys. Rev. Lett.}, 102:021302, 2009.

\bibitem{Carrasco:2012cv}
John Joseph~M. Carrasco, Mark~P. Hertzberg, and Leonardo Senatore.
\newblock {The Effective Field Theory of Cosmological Large Scale Structures}.
\newblock {\em JHEP}, 09:082, 2012.

\bibitem{Baumann:2010tm}
Daniel Baumann, Alberto Nicolis, Leonardo Senatore, and Matias Zaldarriaga.
\newblock {Cosmological Non-Linearities as an Effective Fluid}.
\newblock {\em JCAP}, 07:051, 2012.

\bibitem{Senatore:2017hyk}
Leonardo Senatore and Matias Zaldarriaga.
\newblock {The Effective Field Theory of Large-Scale Structure in the presence of Massive Neutrinos}.
\newblock 7 2017.

\bibitem{Dodelson:2016wal}
Scott Dodelson, Katrin Heitmann, Chris Hirata, Klaus Honscheid, Aaron Roodman, Uro\v{s} Seljak, An\v{z}e Slosar, and Mark Trodden.
\newblock {Cosmic Visions Dark Energy: Science}.
\newblock 4 2016.

\bibitem{LSSTDarkEnergyScience:2018jkl}
Rachel Mandelbaum et~al.
\newblock {The LSST Dark Energy Science Collaboration (DESC) Science Requirements Document}.
\newblock 9 2018.

\bibitem{Kaiser:1991qi}
Nick Kaiser.
\newblock {Weak gravitational lensing of distant galaxies}.
\newblock {\em Astrophys. J.}, 388:272, 1992.

\bibitem{Bartelmann:2016dvf}
Matthias Bartelmann and Matteo Maturi.
\newblock {Weak gravitational lensing}.
\newblock 12 2016.

\bibitem{abell2009lsst}
Paul~A Abell, Julius Allison, Scott~F Anderson, John~R Andrew, J~Roger~P Angel, Lee Armus, David Arnett, SJ~Asztalos, Tim~S Axelrod, Stephen Bailey, et~al.
\newblock Lsst science book, version 2.0.
\newblock 2009.

\bibitem{Yu:2018tem}
Byeonghee Yu, Robert~Z. Knight, Blake~D. Sherwin, Simone Ferraro, Lloyd Knox, and Marcel Schmittfull.
\newblock {Toward neutrino mass from cosmology without optical depth information}.
\newblock {\em Phys. Rev. D}, 107(12):123522, 2023.

\bibitem{Sherwin:2015baa}
Blake~D. Sherwin and Marcel Schmittfull.
\newblock {Delensing the CMB with the Cosmic Infrared Background}.
\newblock {\em Phys. Rev. D}, 92(4):043005, 2015.

\bibitem{Schmittfull:2017ffw}
Marcel Schmittfull and Uros Seljak.
\newblock {Parameter constraints from cross-correlation of CMB lensing with galaxy clustering}.
\newblock {\em Phys. Rev. D}, 97(12):123540, 2018.

\bibitem{Limber:1954zz}
D.~Nelson Limber.
\newblock {The Analysis of Counts of the Extragalactic Nebulae in Terms of a Fluctuating Density Field. II}.
\newblock {\em Astrophys. J.}, 119:655, 1954.

\bibitem{LoVerde:2008re}
Marilena LoVerde and Niayesh Afshordi.
\newblock {Extended Limber Approximation}.
\newblock {\em Phys. Rev. D}, 78:123506, 2008.

\bibitem{Euclid:2023ove}
D.~Sciotti et~al.
\newblock {Euclid preparation - LII. Forecast impact of super-sample covariance on 3\texttimes{}2pt analysis with Euclid}.
\newblock {\em Astron. Astrophys.}, 691:A318, 2024.

\bibitem{DES:2020ypx}
O.~Friedrich et~al.
\newblock {Dark Energy Survey year 3 results: covariance modelling and its impact on parameter estimation and quality of fit}.
\newblock {\em Mon. Not. Roy. Astron. Soc.}, 508(3):3125--3165, 2021.

\bibitem{Ma:2005rc}
Zhao-Ming Ma, Wayne Hu, and Dragan Huterer.
\newblock {Effect of photometric redshift uncertainties on weak lensing tomography}.
\newblock {\em Astrophys. J.}, 636:21--29, 2005.

\bibitem{Shen:2025nat}
Delon Shen, Nickolas Kokron, and Emmanuel Schaan.
\newblock {Direct correlation of line intensity mapping and CMB lensing from lightcone evolution}.
\newblock 7 2025.

\bibitem{Celik:2025wkt}
{\c{S}}afak {\c{C}}elik and Fabian Schmidt.
\newblock {Mixed Dark Matter and Galaxy Clustering: The Importance of Relative Perturbations}.
\newblock 8 2025.

\bibitem{Kaiser:1987qv}
N.~Kaiser.
\newblock {Clustering in real space and in redshift space}.
\newblock {\em Mon. Not. Roy. Astron. Soc.}, 227:1--27, 1987.

\bibitem{LozanoTorres:2022mci}
Jose~Agustin Lozano~Torres and Bjoern~Malte Schaefer.
\newblock {Three-dimensional weak gravitational lensing of the 21-cm radiation background}.
\newblock {\em Mon. Not. Roy. Astron. Soc.}, 512(4):5135--5152, 2022.

\bibitem{Mandel:2005xh}
Kaisey~S. Mandel and Matias Zaldarriaga.
\newblock {Weak gravitational lensing of high-redshift 21 cm power spectra}.
\newblock {\em Astrophys. J.}, 647:719--736, 2006.

\bibitem{Heavens:2003jx}
Alan Heavens.
\newblock {3d weak lensing}.
\newblock {\em Mon. Not. Roy. Astron. Soc.}, 343:1327, 2003.

\bibitem{Zieser:2016lmz}
Britta Zieser and Philipp~M. Merkel.
\newblock {The cross-correlation between 3D cosmic shear and the integrated Sachs{\textendash}Wolfe effect}.
\newblock {\em Mon. Not. Roy. Astron. Soc.}, 459(2):1586--1595, 2016.

\bibitem{Pratten:2016ojf}
Geraint Pratten, Dipak Munshi, Patrick Valageas, and Philippe Brax.
\newblock {3D Weak Lensing: Modified Theories of Gravity}.
\newblock {\em Phys. Rev. D}, 93(10):103524, 2016.

\bibitem{Kitching:2014lga}
T.~D. Kitching, A.~F. Heavens, and S.~Das.
\newblock {3D Weak Gravitational Lensing of the CMB and Galaxies}.
\newblock {\em Mon. Not. Roy. Astron. Soc.}, 449(2):2205--2214, 2015.

\bibitem{CFHTLenS:2014rje}
T.~D. Kitching et~al.
\newblock {3D Cosmic Shear: Cosmology from CFHTLenS}.
\newblock {\em Mon. Not. Roy. Astron. Soc.}, 442(2):1326--1349, 2014.

\bibitem{Xiao:2024qay}
Huangyu Xiao, Liang Dai, and Matthew McQuinn.
\newblock {Detecting dark matter substructures on small scales with fast radio bursts}.
\newblock {\em Phys. Rev. D}, 110(2):023516, 2024.

\bibitem{Mondino:2020rkn}
Cristina Mondino, Anna-Maria Taki, Ken Van~Tilburg, and Neal Weiner.
\newblock {First Results on Dark Matter Substructure from Astrometric Weak Lensing}.
\newblock {\em Phys. Rev. Lett.}, 125(11):111101, 2020.

\bibitem{Wagner-Carena:2024axc}
Sebastian Wagner-Carena, Jaehoon Lee, Jeffrey Pennington, Jelle Aalbers, Simon Birrer, and Risa~H. Wechsler.
\newblock {A Strong Gravitational Lens Is Worth a Thousand Dark Matter Halos: Inference on Small-scale Structure Using Sequential Methods}.
\newblock {\em Astrophys. J.}, 975(2):297, 2024.

\bibitem{Mao:2024pic}
Yao-Yuan Mao et~al.
\newblock {The SAGA Survey. III. A Census of 101 Satellite Systems around Milky Way\textendash{}mass Galaxies}.
\newblock {\em Astrophys. J.}, 976(1):117, 2024.

\bibitem{Nadler:2024ims}
Ethan~O. Nadler, Vera Gluscevic, Trey Driskell, Risa~H. Wechsler, Leonidas~A. Moustakas, Andrew Benson, and Yao-Yuan Mao.
\newblock {Forecasts for Galaxy Formation and Dark Matter Constraints from Dwarf Galaxy Surveys}.
\newblock {\em Astrophys. J.}, 967(1):61, 2024.

\bibitem{Sharma:2025ldt}
Vikhyat Sharma and Arka Banerjee.
\newblock {N-body Simulations of cosmologies with Light Massive Relics}.
\newblock 10 2025.

\bibitem{Leo:2017zff}
Matteo Leo, Carlton~M. Baugh, Baojiu Li, and Silvia Pascoli.
\newblock {The Effect of Thermal Velocities on Structure Formation in N-body Simulations of Warm Dark Matter}.
\newblock {\em JCAP}, 11:017, 2017.

\bibitem{Brandbyge:2017tdc}
Jacob Brandbyge and Steen Hannestad.
\newblock {Cosmological N-body simulations with generic hot dark matter}.
\newblock {\em JCAP}, 10:015, 2017.

\bibitem{Hu:2025lrl}
Rui Hu, Ming-chung Chu, Shek Yeung, and Wangzheng Zhang.
\newblock {Impact of light sterile neutrinos on cosmological large scale structure}.
\newblock {\em JCAP}, 06:014, 2025.

\bibitem{Berlind:2001xk}
Andreas~A. Berlind and David~H. Weinberg.
\newblock {The Halo occupation distribution: Towards an empirical determination of the relation between galaxies and mass}.
\newblock {\em Astrophys. J.}, 575:587--616, 2002.

\bibitem{Mathematica}
Wolfram~Research{,} Inc.
\newblock Mathematica, {V}ersion 14.2.
\newblock Champaign, IL, 2024.

\bibitem{Chen:2020kxi}
Joe~Zhiyu Chen, Amol Upadhye, and Yvonne Y.~Y. Wong.
\newblock {One line to run them all: SuperEasy massive neutrino linear response in $N$-body simulations}.
\newblock {\em JCAP}, 04:078, 2021.

\bibitem{Nascimento:2023psl}
Caio Nascimento.
\newblock {Accurate fluid approximation for massive neutrinos in cosmology}.
\newblock {\em Phys. Rev. D}, 108(2):023505, 2023.

\end{thebibliography}
\bibliographystyle{unsrt}
\end{document}